\documentclass[10pt,journal,compsoc]{IEEEtran}

\usepackage{times}
\usepackage{wrapfig}
\usepackage{epsfig}
\usepackage{amsmath,mathtools,bm}
\usepackage{amssymb}
\usepackage{amsfonts}
\usepackage{graphicx}
\usepackage{caption}
\usepackage{subcaption}
\usepackage{tikz}
\usepackage{tabularx,booktabs}
\usepackage{lipsum}
\usepackage{multirow}
\usepackage{blkarray}
\usepackage{algorithm}
\providecommand{\norm}[1]{\left\lVert#1\right\rVert} %%% Can also add ient-defs-math.tex defenitions

\DeclareOldFontCommand{\bf}{\normalfont\mathbfseries}{\mathbf}

\ifCLASSOPTIONcompsoc

  \usepackage[nocompress]{cite}

\else

  \usepackage{cite}

\fi

\ifCLASSINFOpdf

\else

\fi

\begin{document}
\usetikzlibrary{shapes,arrows}
%






\title{Optimized Feature Space Learning for Generating Efficient Binary Codes for Image Retrieval}

%

%








%

%












\author{Abin~Jose,~\IEEEmembership{Member,~IEEE,}

        Erik Stefan Ottlik,~\IEEEmembership{Member,~IEEE,}

        Christian Rohlfing,~\IEEEmembership{Member,~IEEE,}

        and~Jens-Rainer Ohm,~\IEEEmembership{ Member,~IEEE}% <-this % stops a space

%\IEEEcompsocitemizethanks{\IEEEcompsocthanksitem ----
% \protect\\



%E-mail: ---
%\IEEEcompsocthanksitem --- ---.}% <-this % stops a space
%\thanks{Manuscript received ----}
}




% 



%












\markboth{Journal of \LaTeX\ Class Files,~Vol.~14, No.~8, August~2015}%
{Shell \MakeLowercase{\textit{et al.}}: Bare Advanced Demo of IEEEtran.cls for IEEE Computer Society Journals}



% 

























\IEEEtitleabstractindextext{%
\begin{abstract}
In this paper we propose an approach for learning low dimensional optimized feature space with minimum intra-class variance and maximum inter-class variance. We address the problem of high-dimensionality of feature vectors extracted from neural networks by taking care of the global statistics of feature space. Classical approach of Linear Discriminant Analysis (LDA) is generally used for generating an optimized low dimensional feature space for single-labeled images. Since image retrieval involves both multi-labeled and single-labeled images, we utilize the equivalence between LDA and Canonical Correlation Analysis (CCA) to generate an optimized feature space for single-labeled images and use CCA to generate an optimized feature space for multi-labeled images. Our approach correlates the projections of feature vectors with label vectors in our CCA based network architecture. The neural network minimizes a loss function such that the correlation coefficients are maximized. We binarize our generated feature vectors with the Iterative Quantization (ITQ) approach and also propose an ensemble network to generate binary codes of desired bit length for image retrieval. Our measurement of mean average precision shows competitive results on other state-of-the-art single-labeled and multi-labeled image retrieval datasets.
\end{abstract}
}
\maketitle
\IEEEdisplaynontitleabstractindextext
\IEEEpeerreviewmaketitle
\ifCLASSOPTIONcompsoc
\IEEEraisesectionheading{\section{Introduction}\label{sec:introduction}}
\else
\fi

\IEEEPARstart{I}mages often convey information which words can barely express. If we consider famous paintings, we recognize that we are not capable of exactly describing all our visual impressions. Still, one typically tries to find words to describe the visual content and the higher semantics of images. At present, automatic understanding of images is of high practical relevance. It is used in medicine for disease detection or in autonomous driving for scene understanding \cite{medicalRetrieval2004, auto2018}. Image retrieval can help us to cope with the continuously increasing image datasets in our big data era. 
In image retrieval, we are given a query image, and we search for the most similar images on a vast gallery dataset. The most common approach for image retrieval is to find a feature space where similar images are mapped closely together based on a given metric, e.g., the Euclidean distances \cite{smeulders2000}. Thus, querying for similar images becomes equivalent to a nearest neighbor search in this feature space. Classical methods such as Bag of Visual Words\cite{bow} and Fisher vectors\cite{fv} can generate such a feature space.\par 
These methods can generate a global descriptor which can describe the images but are sometimes incapable of capturing semantic concepts. The problem of capturing higher semantic information is addressed in classification as well.\par In image classification,  AlexNet \cite{alexnet2012} by Krizhevsky et al. outperformed all existing image classification methods significantly leading to the rise of Convolutional Neural Networks (CNNs). CNNs are the basis for all modern state-of-the-art methods in image classification. Furthermore, it was shown that features extracted from CNNs are suitable for image retrieval and outperform classical feature descriptors \cite{babenko2014, siftcnn2018}.
As these features were first extracted from CNNs trained for classification, the CNN's loss did not accurately fit the purpose of image retrieval. In contrast to a standard classification situation, images are often labeled by several concepts in image retrieval. Consequently, a classification loss is unreasonable or cannot be applied here. Therefore, novel methods usually incorporate a pairwise or a triplet loss to generate a good feature space for a retrieval system \cite{dpsh2015}.\par
In content-based image retrieval, we analyze the content of the images to find similar images in a vast database which can contain millions of images \cite{cbirsurvey2008}. The picture we use for the query is referred to as the query image, and the database is called the gallery dataset. Note that for devising the image retrieval methods, one typically expects that for few images labeled data is given. This is necessary to decide which semantics of the images should be captured by the image retrieval method \cite{dtsh2016}. This portion of the dataset will be referred to as the training dataset.
It is important that we need to derive appropriate features suitable for an efficient and effective retrieval system. 
It turns out that feature vectors generated by CNNs provide much better representations for the semantics of the images \cite{ddsh2018}. For instance, the convolutional layers 1-5 of AlexNet derive low-level and mid-level information while the final fully connected layers capture higher semantics of the images. It is necessary to understand which semantics of the pictures should be captured by a retrieval system. As we often cannot describe images by a single concept, it is evident that image retrieval methods have to deal with multi-labeled images as well. While for the simple case of single-labeled images a classification could basically capture the general semantic information, this is not the case for multi-labeled images. For multi-labeled images, we could predict the category labels. This will not lead to a compact feature space suitable for image retrieval, especially, if the number of categories is large as it is difficult to get compact feature space as images contain multiple concepts. The reason why we want to find an appropriate feature space is that finding similar images simplifies to the nearest neighbor search on such a feature space \cite{surveyhash2018}.
Besides the quality of the retrieved results, the efficiency of the retrieval system is a crucial aspect. The challenging sizes of image datasets demand an efficient retrieval system \cite{surveyhash2018}. The typical solution for speeding up the system is to use binary codes instead of the original feature vectors. As a result, we can compute Hamming distances to rank the gallery images
for a query item. The binary codes for the feature vectors shall give an approximation of the original Euclidean distances. The main advantage here is that the Hamming distances are much faster to compute than the Euclidean distances, especially on
modern CPU architectures, Hamming distance computation can be executed faster by an XOR operation followed by a popcnt instruction, i.e., a bit-count.\par
Hashing methods, which provide such a binarization, are therefore indispensable for image retrieval. In hashing, we try to find a hash function that maps our data item to a hash value. Several hash functions are typically used to form the final binary code, which represents an image for the approximate nearest neighbor search.
The hash mapping is in general not injective, i.e., we map different data items to the same hash code \cite{surveyhash2018}. From the image retrieval perspective, we try to map feature vectors to short hash codes while preserving the important similarity information among the original feature vectors. Furthermore, as the probability of collision for similar images is desired to be very high, the similarity-preserving hashing discussed here is different from the traditional hashing methods. Conventional methods typically try to avoid collisions, i.e., they try to map different data items to different hash codes. In image retrieval, we want to assign the similar binary codes for semantically similar images and distinct binary codes for dissimilar images.\par
\subsection{Related Work}
Hashing methods can be divided into data-independent and data-dependent approaches \cite{surveyhash2018,hashsurvey2016}. Data-independent approaches do not fit their hashing functions onto the data. Instead, they have to rely on general objectives and are therefore practically insufficient for an effective image retrieval system. Locality-Sensitive Hashing (LSH) schemes are dominant in this field. They look at the approximate nearest neighbor problem from a probabilistic point of view.
It is natural to wonder if we can find better projections when we know the underlying distribution of our data. Data-dependent hashing methods use the data, i.e., the feature vectors in image retrieval, to find better hash functions. They can be further subdivided into supervised and unsupervised methods. Supervised methods exploit additional information, e.g., label information, to improve the performance of the generated hash codes. On the other hand, unsupervised methods utilize solely the data-distribution with no additional information. Indeed, it turns out that we can find better projections by data-dependent approaches. Thus, research efforts moved to learning-based and data-dependent methods. A well-known unsupervised data-dependent hashing method is spectral hashing proposed by Weiss et al. \cite{sh2009} which is closely related to spectral clustering. They show how hashing projections can be learned from data distributions. 
The binary codes generated by spectral hashing show decent performance, yet especially for a low number of bits, spectral hashing was shown to be inferior to Iterative Quantization (ITQ) \cite{itq2013}. The latter is a simple and very effective unsupervised data dependent binarization method and can be interpreted as fitting an $i$-dimensional hypercube to the $i$-dimensional feature vectors in a way such that the quantization loss is minimized. This approach is computationally efficient and scales well with the data size. Because of its appealing characteristics, it is incorporated in our proposed binarization scheme after learning an optimized feature space.
Recently, supervised deep learning based hashing methods such as Deep Discrete Supervised Hashing (DDSH) \cite{ddsh2018}, Deep Supervised Hashing with Triplet Labels (DTSH) \cite{dtsh2016}, Deep Supervised Hashing with Pairwise Labels (DPSH) \cite{dpsh2015}, Data Sensitive Hashing (DSH) \cite{dsh2016}, Deep Hashing Network (DHN) \cite{dhn2016} have been very successful in image retrieval. Instead of using traditional feature descriptors which are binarized in the second step by classical approaches, deep hashing frameworks usually learn the feature vectors and the binary codes simultaneously. The main problem of the deep hashing approach is that the gradient of the sign-function used for binarization is always zero except at the point zero such that the gradient cannot be calculated to train the CNN end-to-end. Thus, backpropagation cannot be used.
It is noteworthy that unsupervised deep hashing methods exist as well \cite{unsuperviseddeephashing}, \cite{unsup}, \cite{unsup2}, \cite{unsup3}. However, supervised hashing methods are dominant in this field and provide more effective binary codes. The supervised information can be either class label information, pairwise information or triplet labels.
Another example of a successful deep hashing method is HashNet\cite{hashnet2017} which is also often used as a baseline for benchmarking deep hashing frameworks. HashNet uses pairwise similarity information. In the HashNet approach, the slope of a hyperbolic-tangent is reduced till it approximates the sign-function, during training to generate binary codes. However, if one directly starts with a high slope, the network could not train well as the gradient would almost vanish. We will provide benchmarks from our proposed method against this approach which is the current state-of-the-art.
\subsection{This Paper}
In this paper, we propose to generate an optimized feature space which takes care of the global statistics of the feature space such that the between-class scatter is maximum and within-class scatter is minimum to generate a low-dimensional representation space. Linear Discriminant Analysis (LDA) \cite{fisher36lda} is a popular approach to learn an optimized low-dimensional representation space, but is applicable only for single-labeled images. We utilize the relationship between Canonical Correlation Analysis (CCA) \cite{harold1936relations} and LDA and propose a novel feature space learning approach for both single and multi-labeled images. The generated features are binarized using the popular ITQ approach. We also propose a network ensemble method to concatenate the binary codes to generate codewords of desired bit length. We denote the proposed method as Deep Canonical Correlation Feature (DCCF) extraction method in this paper. The retrieval results are measured on standard datasets such as CIFAR-10 \cite{cifar10} which is single-labeled and NUS-WIDE \cite{nuswide} and MS-COCO \cite{mscoco2014} which are multi-labeled datasets. We have compared the performance with other state-of-the-art hashing methods. We also use the feature space visualization using t-SNE \cite{tsne2008} and plot the training curves to show that our feature space learning network's training reaches the lower bound which basically corresponds to the sum of the correlation coefficients of CCA. The major contributions of this paper are summarized below:
\begin{itemize}
   \item Our approach learns an optimized low-dimensional feature space for single-labeled images by utilizing the equivalence between LDA and CCA and addresses the problems of the DeepLDA \cite{deeplda2015} approach (Section \ref{newlda-cca});
   \item We extend the optimized feature space learning to multi-labeled images with CCA approach by designing appropriate category indicator matrices (Section \ref{sec:dclsinglelabeledimages}, \ref{sec:dclimultilabeledimages});
   \item We elaborate from a linear algebra point of view to show how the correlation maximization between the two vectors of CCA can be interpreted for multi-labeled images (Section \ref{sec:cca-la});
   \item We design the loss function based on the correlation coefficients of CCA for training our network to learn the optimized feature space (Section \ref{sec:loss});
   \item We propose an ensemble technique for generating binary codes of desired bit length from the features obtained from our proposed feature extraction technique followed by ITQ (Section \ref{sec:networkensemble}).
\end{itemize}
The paper is organized as follows. Section $2$ explains the theoretical fundamentals of LDA and CCA. Section $3$ explains CCA in detail. In Section $4$, we discuss the proposed feature extraction approach. Section $5$ and $6$ explain how we generate the 
binary codes from the learned low-dimensional feature space. Experimental results are summarized in Section $7$, conclusions and future work are discussed in Section $8$.
%We compare our results with other state-of-the-art hashing approaches such as HashNet \cite{hashnet2017}, DHN (ESR) \cite{dhnesr2016}, DNNH \cite{dnnh2015}, CNNH \cite{cnnh2014}, SDH \cite{sdh2015}, KSH \cite{ksh2012}, ITQ-CCA \cite{itq2013}, ITQ \cite{itq2013}, BRE \cite{bre2009}, SH  \cite{sh2009}, and LSH \cite{lsh1999}.  HashNet, DHN (ESR), DNNH, and CNNH are deep hashing based methods. SDH, KSH, and ITQ-CCA are non-deep supervised hashing methods. ITQ, BRE, and SH are unsupervised methods and LSH is a data-independent method  
%(Write one para about contibutions todo -Abin)%\section{Linear Discriminant Analysis} \label{sec:FUNLDA}
\vspace{0.1mm}
\section{PRELIMINARIES}
In this Section, we explain how Linear Discriminant Analysis (LDA) \cite{fisher36lda} and Canonical Correlation Analysis (CCA) \cite{harold1936relations} can generate a low dimensional feature space. We also explain the equivalence between LDA and CCA. We also briefly introduce the neural network based approach for learning LDA, called DeepLDA \cite{deeplda2015}. This approach is not efficient for retrieval in case of multi-labeled images as DeepLDA learns the LDA projection which is the dimensionality reduction for single-labeled images. We address this problem by using CCA for correlating the projections between the feature vectors and label vectors. For single-labeled images, we utilize the equivalence between LDA and CCA and find the $\mathrm{C-1}$ correlation coefficients, where $\mathrm{C}$ is the number of classes. For multi-labeled images,
we project to the low-dimensional\footnote{Due to memory restrictions of the GPU, we cannot ensure all categories are present in the training batches. Therefore, we project to a low dimensional space $< \mathrm{C}$ as explained in Subsection \ref{nuswide-final}.} space $\mathrm{C}$.
\subsection{Linear Discriminant Analysis} \label{sec:FUNLDA}
In Section 1, we have already mentioned briefly about deep hashing as a possibility for generating highly efficient and effective binary codes for image retrieval. Deep hashing networks learn effective feature representations internally. The loss functions applied are often built on pairwise, or triplet information, so it is questionable if they can reflect the global statistics of the data. This motivated us for designing an LDA based retrieval system. LDA is a method for dimensionality reduction such that the low-dimensional feature space could have high inter-class variance and small intra-class variance. LDA observes the global structure of the data by estimating covariance matrices \cite{fisherlda2005,deeplda2015}. Fisher proposed this approach for two class problem and learned the projection matrix which separates the two classes by maximizing the ratio of the between-class scatter to the within-class scatter \cite{fisherlda2005}. The between-class scatter matrix $\mathbf{S}_\mathrm{B}$ and the within-class scatter matrix $\mathbf{S}_\mathrm{W}$ for a general multi-class problem is defined as:
%\begin{equation} \notag
%\mathbf{S}_W = \sum_{c=1}^C \sum_{i=1}^{n_c} (\mathbf{x}_{ci} - \bar{\mathbf{x}}_c) (\mathbf{x}_{ci} - \bar{\mathbf{x}}_c)^T
%\end{equation}
%\begin{equation} \notag
%\mathbf{S}_B = \sum_{c=1}^C n_c \bar{\mathbf{x}}_c \bar{\mathbf{x}}_c^T
%\end{equation}
\begin{equation} \notag %\label{eq:thematricesforlda}
\begin{split}
& \mathbf{S}_\mathrm{B} = \sum_{c=1}^\mathrm{C} \frac{n_c}{n} ( \bar{\mathbf{x}}_c - \bar {\mathbf{x}} ) ( \bar{\mathbf{x}}_c - \bar {\mathbf{x}} )^T,\\
& \mathbf{S}_\mathrm{W} = \sum_{c=1}^\mathrm{C} \frac{1}{n} \sum_{i=1}^{n_c}  (\mathbf{x}_{ci} - \bar{\mathbf{x}}_c) (\mathbf{x}_{ci} - \bar{\mathbf{x}}_c)^T,
\end{split}
\end{equation}
where $\mathrm{C}$ denotes the number of classes, $\mathbf{x}_{ci}$ is the $i^{\mathrm{th}}$ feature vector of class $c$, $n_c$ is the number of images belonging to class $c$, and $\bar{\mathbf{x}}_c$ is the mean of the feature vectors belonging to class $c$. The total number of images is denoted as $n$ and $\bar{\mathbf{x}}$ represents the global mean.
LDA finds the projection matrix $\mathbf{A}^* \in \mathbb{R}^{{t} \times {i}},$ such that the objective
\begin{equation} \label{ldaequ}
\mathbf{A}^* = \mathrm{arg\max\limits_\mathbf{A}} \left(\frac{ \vert \mathbf{A} \mathbf{S}_\mathrm{B} \mathbf{A}^T \vert }{ \vert \mathbf{A} \mathbf{S}_\mathrm{W} \mathbf{A}^T \vert} \right) 
\end{equation}
is maximized. Here $t$ denotes the high-dimensional feature vector and $i$ corresponds to the LDA dimensions. Maximizing this ratio generates the matrix $\mathbf{A}^*$ which consists of those eigenvectors $\mathbf{a}_i$ of the matrix $\mathrm{\mathbf{S}_W^{-1}} \mathbf{S}_\mathrm{B}$ corresponding to the largest eigenvalues. The number of discriminant directions, $i$ is at most $\mathrm{C-1}$ which is the rank of $\mathrm{\mathbf{S}_W^{-1} \mathbf{S}_B}$.
Thus, we can only project to a space with maximum $\mathrm{C-1}$ dimensions \cite{mardia}.\par The very first attempt in learning the LDA projection using neural networks, DeepLDA was proposed by Dorfer et al. \cite{deeplda2015} in 2016. DeepLDA applies LDA-based loss to train a CNN end-to-end. By doing this, a feature space can be found which is suitable for classification. The authors propose to maximize the eigenvalues in all $\mathrm{C-1}$ eigenvector directions such that discrimination is achieved in all directions. Moreover, all the classes have to be mutually exclusive. In the context of image retrieval, this means that only single-labeled images could be used when DeepLDA is applied for the feature extraction. However, with multi-labeled images, we could use a variant of LDA called CCA. Generating an optimized feature space for multi-labeled images is the main problem we are addressing in this paper. For that, we introduce the concept of CCA next.
\subsection{Canonical Correlation Analysis}
Hotelling first proposed Canonical Correlation Analysis (CCA) in 1936 \cite{harold1936relations}. It is a very general concept in statistics, and many tests in statistics can be regarded as special forms of CCA. CCA can be viewed as a generalization of multiple regression \cite{mardia}, where we want to find linear projections to maximally correlate data vectors to a one-dimensional predictor variable. In CCA, this is extended such that we maximize the correlation between the data vectors and vectors from a second dataset. Given two datasets, we try to find the best linear projection such that the correlation in the projected space for the two data vectors is maximized. To describe CCA, we start with two column vectors representing the random variables for the two data vectors, $\mathbf{X}$ and $\mathbf{Y}$ as:
\begin{equation} \notag
\mathbf{X} = \begin{pmatrix} \mathbf{x}_1 \\ \vdots \\ \mathbf{x}_q \end{pmatrix} , \quad \mathbf{Y} = \begin{pmatrix} \mathbf{y}_1 \\ \vdots \\ \mathbf{y}_{q} \end{pmatrix},
\end{equation}
where $q$ denotes the number of random variables, which will be the batchsize in context of neural networks.
In CCA, we want to find linear projections of the two data vectors,  such that the correlation, $\mathrm{corr}(\mathbf{a}^T \mathbf{X}, \mathbf{b}^T \mathbf{Y})$ is maximized which can be expressed as:  
\begin{equation} 
\begin{split}
\rho \hspace{.1cm} \left( \mathbf{a^*},\mathbf{b^*} \right) &=  \max\limits_{\mathbf{a},\mathbf{b}} \mathrm{corr} (\mathbf{a}^T \mathbf{X}, \mathbf{b}^T \mathbf{Y}), 
\label{eq:corr}
\end{split}
\end{equation}
where $\mathrm{\rho}$ is the correlation coefficient and $\mathbf{a^*}$ and $\mathbf{b^*}$ are the optimal projections.
\subsection{Equivalence between Linear Discriminant Analysis and Canonical Correlation Analysis} \label{newlda-cca}
Maurice Bartlett first proved the equivalence between LDA and CCA in 1938 \cite{bartlett1938}. In our paper, exactly this relationship is used to generate an optimum feature space using CNN architecture. Bartlett precisely described the relationship between the eigenvalues ($\mathrm{\lambda_{LDA}}$) of the LDA and the eigenvalues ($\mathrm{\lambda_{CCA}}$) of the CCA as:
\begin{equation} \label{New-test}
\begin{split}
& \mathrm{\lambda_{CCA}} = \frac{\mathrm{\lambda_{LDA}}}{1 + \mathrm{\lambda_{LDA}}} \cdot \\
\end{split}
%\label{New-test}
\end{equation}
Knowing this exact relationship between the eigenvalues is useful for designing the loss function for a network generating an optimized feature space as described in Subsection \ref{sec:loss}. A problem of the DeepLDA \cite{deeplda2015} approach is that the eigenvalues are not bounded. Hence, the network tends to produce trivial solutions when the negative sum of all eigenvalues is used as the loss as proposed in the DeepLDA approach. A network trained with such an objective tries to push one eigenvalue to a very large value, while the other eigenvalues stay close to zero. This is basically the trivial solution of LDA. This means that only already well separated classes are pushed further apart in the feature space.\par From \eqref{New-test}, we can see that the eigenvalues of CCA are upper bounded by one ($\mathrm{\lambda_{CCA}} \leq 1$). Therefore, a network trained with an objective built upon the CCA eigenvalues cannot push the eigenvalue of one dimension to an arbitrarily large value since it will never get larger than one. Instead, once a dimension is well separated, i.e., if the corresponding CCA eigenvalue is close to one, the eigenvalues of the other dimensions have to be increased by the network for minimizing the loss. This is due to the fact that the loss function is designed as the sum of the correlation coefficients and the network tries to maximize this correlation in all the discriminant dimensions (cf. Sec. \ref{sec:loss}). This is a significant advantage of the CCA based loss function in a CNN architecture, because the network will generate a feature space with a good separation in all dimensions.\par

\section{Generation of data vectors of Canonical Correlation Analysis} \label{sec:ccaldaequivalence} % 3 pages
In this section, we explain how we can generate the two data vectors of CCA for the case of single-labeled and multi-labeled images.
%We follow their notation, yet we deviate at a later point in this proof from their given proof. For following their notation which simplifies the proof, we slightly change the notation of the between-class scatter matrix $\mathbf{S}_B$ and the within-class scatter matrix $\mathbf{S}_W$. According to Bach et al., we first define $\pi_i = n_i / n$ where $n_i$ is the number of data points belonging to class $i$ and $n = \sum_i n_i$ \cite{bach2005}. The number of data points will be later our batch size during the training of the CNN. From the means of the different classes, we construct the data matrix $\mathbf{M} = (\mathbf{m}_1, \dots, \mathbf{m}_k)$ where $\mathbf{m}_i$ denotes the mean vector of our feature vectors corresponding to class $i$ and $k$ is the number of classes \cite{bach2005}. We can then define the within-class covariance matrix and the between-class covariance matrix \cite{bach2005}:
\subsection{Canonical Correlation Analysis for Single-labeled Images} \label{sec:dclsinglelabeledimages}
First we discuss how we generate the two data vectors for single-labeled images. We use the notation $\mathbf{x}$ $\in$ $\mathbb{R}^{\mathrm{C}}$ for an image feature vector, where $\mathrm{C}$ is the number of classes. Matrix $\mathbf{X} \in \mathbb{R}^{q\times \mathrm{C}}$ contains the feature vectors $\mathbf{x}$ as the rows, where $q$ denotes the batchsize. Now we consider the case, where we construct the second data vector of CCA, i.e., the label vector $\mathbf{y} \in \mathbb{R}^{\mathrm{C}}$, where $\mathrm{C}$ denotes the number of classes, by:
\begin{equation} \notag
\mathbf{y}_i =
\begin{cases}
1, \text{ if image belongs to class i}\\
0, \text{ otherwise}.\\
\end{cases}
\end{equation}
We will denote the matrix $\mathbf{Y} \in \mathbb{R}^{q \times \mathrm{C}}$ as the matrix which contains the sampled versions of the label vectors $\mathbf{y}$ as its rows. For our training images, we have the label information given such that we can construct the second data vector $\mathbf{Y}$ as:
\begin{equation} \notag
\mathbf{Y}_{ij} =
\begin{cases}
1, \text{ if image i belongs to class j}\\
0, \text{ otherwise}.\\
\end{cases}
\end{equation}
We call this second data vector as category indicator matrix.
\subsection{Canonical Correlation Analysis for Multi-labeled Images} \label{sec:dclimultilabeledimages}
\begin{figure}
\begin{center}
  \includegraphics[height = 0.4\textwidth , width = 0.45\textwidth]{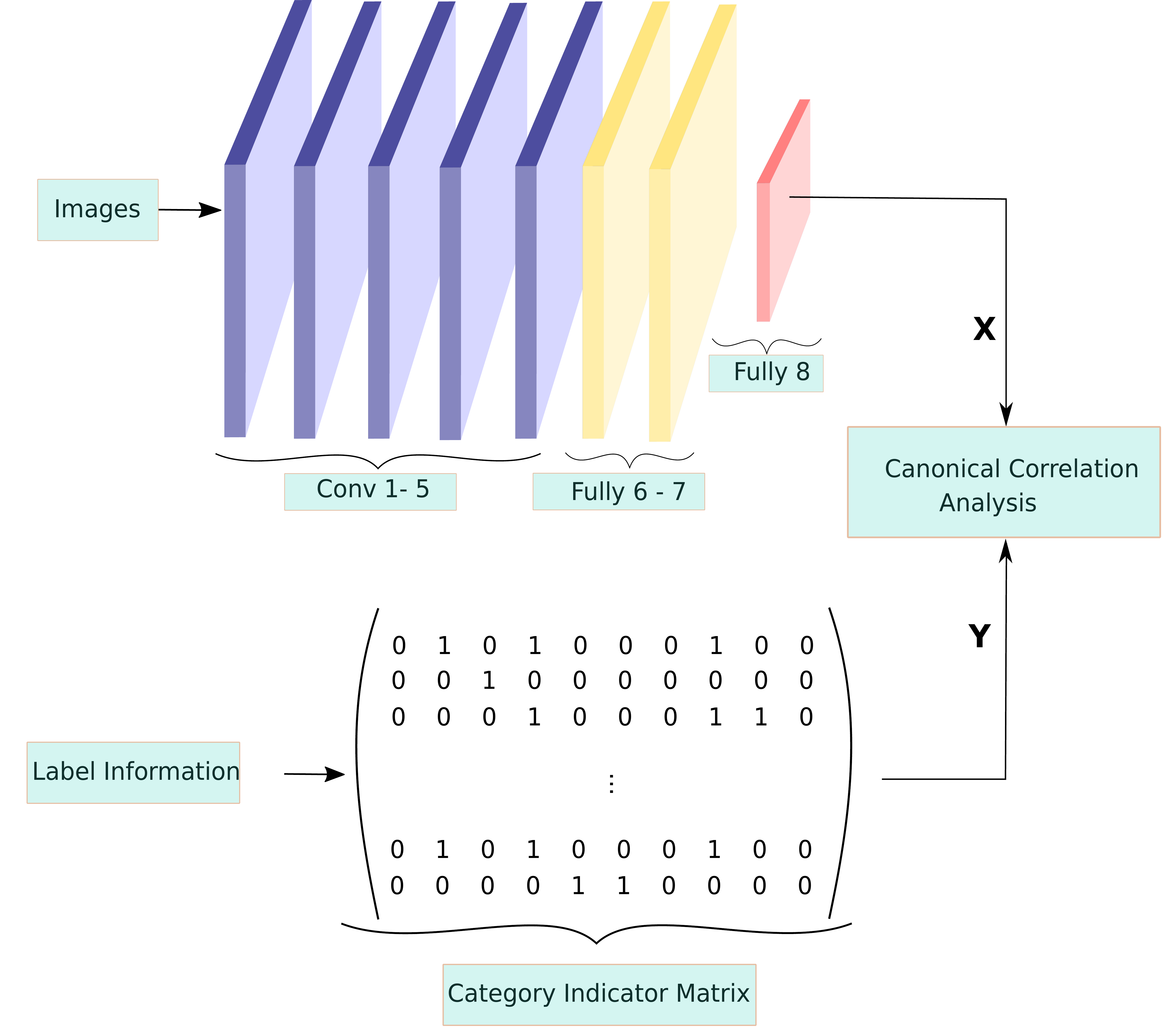}
  \captionsetup{justification=centering}
  \caption{\scriptsize Deep Canonical Correlation Feature (DCCF) extraction approach. The first data vector $\mathbf{X}$ is generated from feature vectors and the second data vector $\mathbf{Y}$ from labels. The CCA is then computed on this random variables. The network is finetuned using the loss function \eqref{eq:ourloss}. }
  \label{fig:deepccalda}
\end{center}
\end{figure}

In this subsection, we explain how we can generate the two data vectors for multi-labeled images. By using a category indicator matrix as the second data vector and by utilizing the relationship \eqref{New-test}, we can have the same excellent properties of LDA by applying CCA for single-labeled images. The problem of DeepLDA approach for feature extraction is that it cannot be applied to multi-labeled images since the classes are not mutually exclusive. Here comes the advantage of CCA as it can be easily extended to multi-labeled images. We use the straight-forward extension by modifying the category indicator matrix. A modified category indicator matrix $\mathbf{Y}$ is as shown in Fig. \ref{fig:deepccalda}. Thus, the rows of this matrix will contain a one indicating the corresponding category of that particular image.  
% \[
% \footnotesize
% \begin{blockarray}{cccccc}
% &  & \mathrm{class} \: {1} & \mathrm{class} \: {2} & \mathrm{class} \: {3} & \mathrm{class} \: {4}\\
% \begin{block}{cc(cccc)}
%    & \mathrm{image}\: 1 \textrm{  } & 1 & 0 & 1 & 0 \\
%   \mathbf{Y}_{\mathrm{multi}} =  & \mathrm{image}\: 2 \textrm{  } & 0 & 1 & 0 & 1\\
%    & \vdots & \vdots & \vdots & \vdots \\
%    & \mathrm{image}\: {b} \textrm{  } & 1 & 0 & 0 & 1\\
% \end{block}
% \end{blockarray}.
%  \]
The output of our CNN, $\mathbf{X}$ is still used as the first data vector for CCA in case of multi-labeled images. Thus, we could easily extend the CCA approach to the CNN architecture for multi-labeled images to generate a good feature space.\par
\subsection{Canonical Correlation Analysis - Linear Algebra Point of View} \label{sec:cca-la}
\begin{figure}
\begin{center}
  \includegraphics[scale=0.4, angle=0]{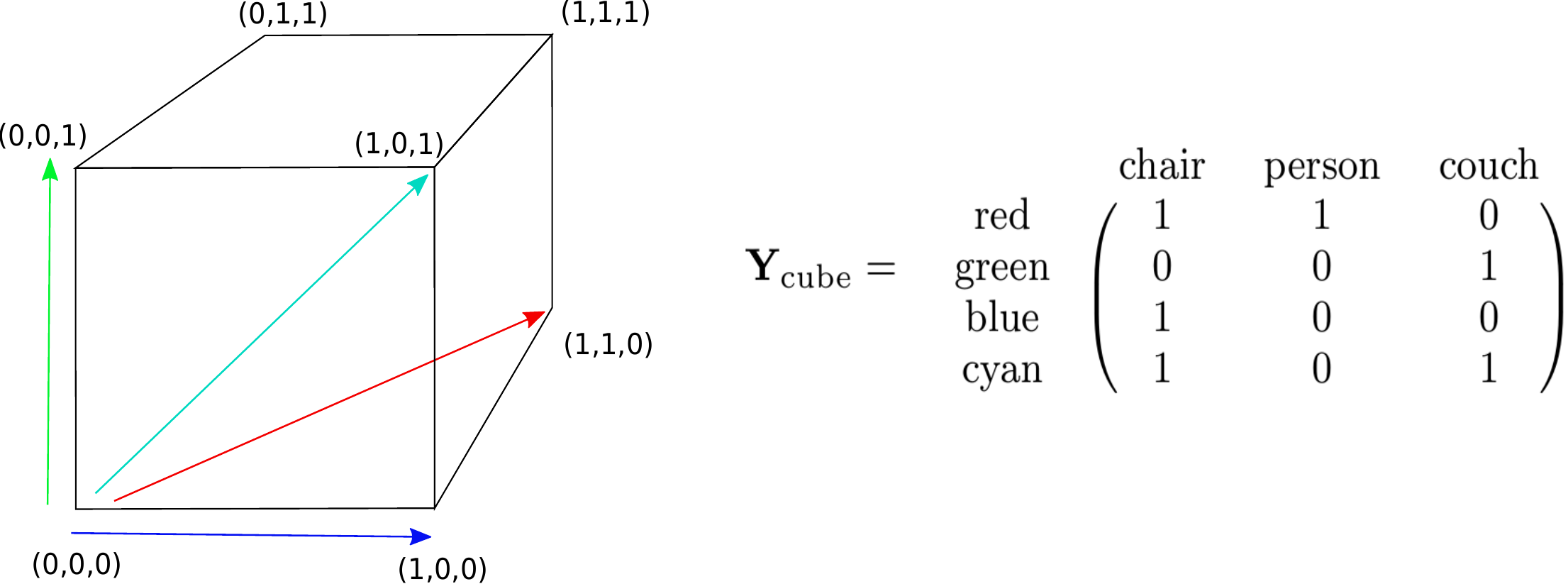}   
  %\captionsetup{justification=centering}
  \caption{\scriptsize Correlation visualization. Each label vector corresponds to a row of the category indicator matrix $\mathbf{Y}_{\mathrm{cube}}$. In CCA, the projected network outputs point in the same direction of the label vectors minimizing the angle between the two and thus generating a feature space with maximum separation between the feature vectors.}
  \label{fig:multilabeledCorrVis}
\end{center}
\end{figure}

Now we elaborate, how the correlation between the two data vectors which is maximized by CCA can also be interpreted from a linear algebra point of view. It can be easily derived that the covariance satisfies the properties of an inner product in a vector space \cite{innerproduct2011}. For the random variables $\mathbf{U}$ and $\mathbf{V}$, 
the angle between two vectors is given by:
\begin{equation} \notag
\mathrm{cos} (\mathrm\theta) = \frac{ \langle \mathbf{X}, \mathbf{Y} \rangle } { \norm{\mathbf{X}} \norm{\mathbf{Y}} } = \frac{\mathrm{cov}(\mathbf{X},\mathbf{Y})}{\sigma_\mathbf{X} \sigma_\mathbf{Y}} = \mathrm{corr}(\mathbf{X},\mathbf{Y}) = \rho,
\end{equation}
where corr indicates the correlation and cov indicates the covariance and the standard deviation of $\mathbf{X}$ and $\mathbf{Y}$  are denoted by $\sigma_\mathbf{X}$ $\sigma_\mathbf{Y}$.
We see that the correlation coefficient $\rho$ is exactly the cosine of the angle between the two random variables. Note that the linear projections found by CCA essentially project to new random variables which can be written as $\mathbf{X}$ and $\mathbf{Y}$. Therefore, we can interpret CCA in a way that we are trying to minimize the angle between our first data vector and our second data vector in the projected space. This explains why our extension to multi-labeled images is meaningful. In Fig. \ref{fig:multilabeledCorrVis}, a cube for 3-dimensional feature vectors is depicted. 
In this case, the different colors correspond to images depicting the concepts listed in the columns of the matrix. The rows of this matrix are essentially label vectors which are shown in Fig. \ref{fig:multilabeledCorrVis}. The label vectors point to the different corners of the cube.\par From the linear algebra point of view, it makes sense to apply CCA for multi-labeled images as we will minimize the angles between the projected network output and the projected label vectors by maximizing the correlation. When the feature vectors from the network output point in the same direction as the label vectors, we have a well-separated feature space.
The intuition with the given cube, also explains why it is reasonable to combine our feature space generation with Iterative Quantization (ITQ) for the final binarization part. This combination and the specific reason for choosing ITQ will be explained in Section \ref{sec:combinationwithitq} in detail.

\section{Proposed feature extraction technique} \label{sec:DeepCCALDA} % 2 pages
Now, we explain in detail the framework of Deep Canonical Correlation Feature (DCCF) extraction.
We have already explained the equivalence of LDA to CCA in Subsection \ref{newlda-cca} and explained how the two data vectors for CCA can be generated for single-labeled and multi-labeled images. Now, we will show how we can apply CCA in a neural network architecture by incorporating the category indicator matrix as the second data vector. Andrew et al.\cite{andrew2013} have already used CCA in a neural network architecture. They use their framework to correlate speech and articulatory data. We correlate the output of a CNN which was given an image as input with specially designed category indicator matrices as the second data vector. The structure of our framework is depicted in Fig. \ref{fig:deepccalda}. The first data vector, $\mathbf{X}$ is obtained from the final layer of the neural network and the second data vector, $\mathbf{Y}$ is formed from the category indicator matrix and the correlation between the two data vectors are computed. The loss function $\L$, with which the network is trained is explained in the next subsection. With our proposed method, we can exploit the equivalence of CCA to LDA for single-labeled images and make it simultaneously applicable to multi-labeled images. Eventually, we can find an optimized feature space by this procedure which can be easily binarized by ITQ to obtain binary codes.
\subsection{Design of Loss Function for Neural Network} \label{sec:loss}
The critical aspect for training our network end-to-end is the loss computation. To discuss the derivation of the loss function, we first present how we can solve CCA as a generalized eigenvalue problem, which could theoretically be implemented as the loss function for training the CNN architecture. For solving CCA, we want to find the optimal projections $\mathbf{a^*}$ and $\mathbf{b^*}$ which maximize the correlation coefficients $\rho$ in \eqref{eq:corr}.
% \begin{equation} \notag
% %    \rho(\mathbf{a^{*}},\mathbf{b^{*}}) = \mathrm{arg \max\limits_{\mathbf{a},\mathbf{b}} }\frac{\mathrm{Cov}(\mathbf{a}^T\mathbf{X},\mathbf{b}^T\mathbf{Y})}{\sigma_{\mathbf{a}^T\mathbf{X}} \sigma_{\mathbf{b}^T\mathbf{Y}}}
% \end{equation}
First we define the sampled covariance matrices $\mathbf{\Sigma}_{11} = \mathrm{cov}(\mathbf{X},\mathbf{X})$, $\mathbf{\Sigma}_{12} = \mathrm{cov}(\mathbf{X},\mathbf{Y})$, 
and $\mathbf{\Sigma}_{22} = \mathrm{cov}(\mathbf{Y},\mathbf{Y})$. The correlation coefficient $\rho$ in \eqref{eq:corr} can be expressed as:
\begin{equation} 
\rho \hspace{.1cm} \left (\mathbf{a^{*}},\mathbf{b^{*}} \right) = \max\limits_{\mathbf{a},\mathbf{b}} \frac{\mathbf{a}^T \mathbf{\Sigma}_{12} \mathbf{b}}{\sqrt{\mathbf{a}^T \mathbf{\Sigma}_{11} \mathbf{a} \mathbf{b}^T \mathbf{\Sigma}_{22} \mathbf{b} } } \cdot
\label{equ:rho}
\end{equation}
$\rho$ is invariant to rescaling the projections $\mathbf{a}$ and $\mathbf{b}$ \cite{hardoon2004}. Therefore, we can constrain them such that $\mathbf{a}^T \mathbf{\Sigma}_{11} \mathbf{a} = 1$ and $\mathbf{b}^T \mathbf{\Sigma}_{22} \mathbf{b} = 1$.
Mardia et al. \cite{mardia} have already derived the eigenvalue solution of CCA by using Singular Value Decomposition (SVD) which is briefly summarized here. First, we define the matrices $\mathbf{K}$, $\mathbf{N_{1}}$, $\mathbf{N_{2}}$, $\mathbf{M_{1}}$ and $\mathbf{M_{2}}$ as:
\begin{equation} \label{matrices}
\begin{split}
& \mathbf{K}   \hspace{.35cm} =  \hspace{.3cm} \mathbf{\Sigma}_{11}^{-1/2} \mathbf{\Sigma}_{12} \mathbf{\Sigma}_{22}^{-1/2}, \\
& \mathbf{N}_1 \hspace{.23cm}  =  \hspace{.23cm} \mathbf{K} \mathbf{K}^T, \\
& \mathbf{N}_2 \hspace{.23cm}  = \hspace{.23cm} \mathbf{K}^T \mathbf{K}, \\
&\mathbf{M}_1  \hspace{.2cm}  = \hspace{.2cm} \mathbf{\Sigma}_{11}^{-1/2} \mathbf{N}_1 \mathbf{\Sigma}_{11}^{1/2} = \mathbf{\Sigma}_{11}^{-1} \mathbf{\Sigma}_{12} \mathbf{\Sigma}_{22}^{-1} \mathbf{\Sigma}_{21}, \\
&\mathbf{M}_2  \hspace{.2cm}  = \hspace{.2cm} \mathbf{\Sigma}_{22}^{-1/2} \mathbf{N}_2 \mathbf{\Sigma}_{22}^{1/2} = \mathbf{\Sigma}_{22}^{-1} \mathbf{\Sigma}_{21} \mathbf{\Sigma}_{11}^{-1} \mathbf{\Sigma}_{12}. \\
\end{split}
\end{equation}\par
The eigenvectors of the matrix $\mathbf{M}_1$ are the solutions of the optimal linear projections $\mathbf{a}^{*}_{i}$, where $i$ is the number of CCA dimensions. It is easy to see that the eigenvectors of $\mathbf{M}_2$ are the solutions for the optimal $\mathbf{b}^{*}_{i}$ due to the symmetry of CCA, i.e., we can interchange the two data vectors. Now, we deviate from the proof in \cite{mardia} and directly show that if $\mathbf{\bm{\alpha}}$ is an eigenvector of $\mathbf{N}_1$, then $\mathbf{a} = \mathbf{\Sigma}_{11}^{-1/2} \mathbf{\bm{\alpha}}$ is an eigenvector of $\mathbf{M}_1$ as shown below:
\begin{equation} \label{eq:svdSameEig}
\begin{split}
& \mathbf{N}_1 \mathbf{\bm{\alpha}} = \lambda_{\mathrm{\scriptsize CCA}} \mathbf{\bm{\alpha}}, \\
& \mathbf{\Sigma}_{11}^{-1/2} \mathbf{N}_1 \mathbf{\bm{\alpha}} = \lambda_{\mathrm{\scriptsize CCA}} \mathbf{\Sigma}_{11}^{-1/2} \mathbf{\bm{\alpha}}, \\
& \mathbf{\Sigma}_{11}^{-1/2} \mathbf{N}_1 \mathbf{\Sigma}_{11}^{1/2} \mathbf{a} = \lambda_{\mathrm{\scriptsize CCA}} \mathbf{a}, \\
& \mathbf{M}_1 \mathbf{a} = \lambda_{\mathrm{\scriptsize CCA}} \mathbf{a}.
\end{split}
%\label{eigens}
\end{equation}\par
Thus, if we find the eigenvectors $\mathbf{\bm{\alpha}}$ of $\mathbf{N}_1$ we can easily compute the eigenvectors $\mathbf{a} = \mathbf{\Sigma_{11}}^{-1/2} \bm{\alpha}$ of $\mathbf{M}_1$ which are the solutions for the optimal linear projections for the first data vector. The same holds for $\mathbf{b} = \mathbf{\Sigma}_{22}^{-1/2} \mathbf{\bm{\beta}}$ due to the symmetry of CCA, where $\bm{\beta}$ is the corresponding eigenvector of $\mathbf{N}_2$. The important aspect is that we can easily compute the eigenvectors of $\mathbf{N}_1$ by an SVD of $\mathbf{K}$ as:
\begin{equation} \notag
\mathbf{K} = (\mathbf{\bm{\alpha}}_1, \dots, \mathbf{\bm{\alpha}}_k) \mathrm{diag}(\bm{\sigma}) (\mathbf{\bm{\beta}}_1, \dots, \mathbf{\bm{\beta}}_k)^T ,
\end{equation}
where $k$ is the rank of the matrix.
The diagonal matrix is of special interest as this one reveals the maximum value of the objective \eqref{equ:rho} for the optimal linear projections $\mathbf{a}^{*}_{i}$ and $\mathbf{b}^{*}_{i}$ since:
\begin{equation} \notag
\begin{split}
& {\mathbf{a}^{*}_{i}}^T \mathbf{\Sigma_{12}} \mathbf{b}^{*}_{i} \\
& = {\mathbf{a}^{*}_{i}}^T \mathbf{\Sigma}_{11}^{1/2} \mathbf{K} \mathbf{\Sigma}_{22}^{1/2} \mathbf{b}^{*}_{i} \\
& = (\mathbf{\Sigma}_{11}^{-1/2} \mathbf{\bm{\alpha}}_i)^T \mathbf{\Sigma}_{11}^{1/2} \mathbf{K} \mathbf{\Sigma}_{22}^{1/2} \mathbf{\Sigma}_{22}^{-1/2} \mathbf{\bm{\beta}}_i \\
& = \mathbf{\bm{\alpha}}_i^T (\mathbf{\bm{\alpha}}_1, \dots, \mathbf{\bm{\alpha}}_k) \mathrm{diag}(\bm{\sigma}) (\mathbf{\bm{\beta}}_1, \dots, \mathbf{\bm{\beta}}_k)^T \mathbf{\bm{\beta}}_i\\
& = \sigma_i,
\end{split}
\end{equation}
where $i$ denotes the number of correlation coefficeints of CCA.\par
Thus, the singular value $\sigma_i$ will be the result of the maximization for the optimal linear projections $\mathbf{a}^{*}_{i}$ and $\mathbf{b}^{*}_{i}$. This means that the singular values obtained are exactly the maximum correlation coefficients and therefore bounded by one ($\sigma_i \leq 1$). The correlation coefficients are generally bounded by $-1 \leq \rho \leq 1$.
This can also be inferred from the linear algebra point of view as the cosine of the angle between two projected random variables and it has the same limits (cf. Sec. \ref{sec:cca-la}).
From \eqref{eq:svdSameEig}, we know that the eigenvalues of $\mathbf{N}_1$ and $\mathbf{M}_1$ are the same, $\lambda_{\mathrm{CCA}}$. This means that the squared singular values derived by the SVD of $\mathbf{K}$ are also the eigenvalues of $\mathbf{M}_1$. By using the relationship in \eqref{New-test}, we can express the canonical correlation coefficients in terms of the eigenvalues found by CCA for mutually-exclusive classes by:
\begin{equation} \label{final_equation}
\rho_i = \sigma_i = \sqrt{{\lambda_{\mathrm{CCA}i}}} = \sqrt{ \frac{{\lambda_{\mathrm{LDA}i}}}{1+ {\lambda_{\mathrm{LDA}i}}} } \cdot
\end{equation}\par
Therefore, to compute the loss for the proposed neural network framework, we first calculate the matrix $\mathbf{K}$ in \eqref{matrices}. For the computation of $\mathbf{K}$, the root inverse matrices of the sampled covariance matrices are necessary. They can be computed utilizing a symmetric eigenvalue decomposition on the sampled covariance matrices. Then, we compute the SVD of $\mathbf{K}$ to compute the singular values $\sigma_i$ which correspond to the correlation coefficients $\rho_i$. Finally, we can return the negative sum of the singular values as the loss. This is because, the neural network loss function has to be minimized and hence we take the negative of the sum. The network targets to maximize the correlation coefficients as maximizing the correlation coefficients will align projected network output and projected label vectors. Therefore, at the end of training the singular values, equivalently the correlation coefficients are maximized. The loss $\L$ which should be minimized to maximize the correlation coefficients is:
\begin{equation} \label{eq:ourloss}
\begin{split}
\L   &= - \sum_{i=1}^{k} \sigma_i = - \sum_{i=1}^{k} \rho_i \\
& \stackrel{\mathclap{\normalfont\mbox{\tiny for LDA equiv.}}}{=} \quad -\sum_{i=1}^k \sqrt{ \frac{{\lambda_{\mathrm{LDA}i}}}{1+ {\lambda_{\mathrm{LDA}i}}} } \cdot
\end{split}
\end{equation}
The algorithm for feature space learning using the proposed DCCF method is summarized in Algorithm \ref{algo:b}.
%This is a advantage of this formulation. Incorporation of CCA loss in the CNN architecture is summarized in Algorithm \ref{algo:b}.
%The singular values which we use for the loss are also present in several matrix norms of the computed matrix $\mathbf{K}$ which leads to different loss variations which we will discuss next.
%The singular values computed by the SVD represent the correlations. While we try to maximize the eigenvalues in case of DeepLDA in Deep CCA these singular values are maximized to increase the correlation. The singular values of the T matrix is Ky-Fan norm of order lof the T matrix. If we choose all singular values we can approximate the sum of the singular values by the blabla norm. In that case, we actually maximize the sum of the square roots of the singular values.
\begin{algorithm}
\caption{. DCCF Loss Calculation}
%\DontPrintSemicolon % Some LaTeX compilers require you to use \dontprintsemicolon    instead
 \textbf{Input:} {\\
a. CNN feature vectors $\mathbf{X}$ of the current batch\\ 
b. Category indicator matrix $\mathbf{Y}$ \\ 
c. Number of correlation coefficients $i$}\\
  \textbf{Output:}\\ {a. Loss value $\L$}\\
1. Calculate the sampled covariance matrices: \\
  \hspace{10cm}$\mathbf{\Sigma}_{11} = \mathrm{cov}(\mathbf{X},\mathbf{X})$ ;  $\mathbf{\Sigma}_{12} = \mathrm{cov}(\mathbf{X},\mathbf{Y})$  ; $\mathbf{\Sigma}_{22} = \mathrm{cov}(\mathbf{Y},\mathbf{Y})$. \\
2. Calculate the symmetric eigenvalue decomposition of : \\
  \hspace{10mm}$\mathbf{\Sigma}_{11} = \mathbf{V}_{1} \mathrm{diag}({\mathbf{\lambda}_{1}}) \mathbf{V}_{1}^T$ ; \hspace{1mm}$\mathbf{\Sigma}_{22} = \mathbf{V}_{2} \mathrm{diag}({\mathbf{\lambda}_{2}}) \mathbf{V}_{2}^T$.\\
3. Calculate the root inverse: \\
 \hspace{10mm}$\mathbf{\Sigma}_{11}^{-1/2} = \mathbf{V}_1 \mathrm{diag}({\mathbf{\lambda}_{1}^{-1/2}}) \mathbf{V}_{1}^T$; $\mathbf{\Sigma}_{22}^{-1/2} = \mathbf{V}_2 \mathrm{diag}({\mathbf{\lambda}_{2}^{-1/2}}) \mathbf{V}_{2}^T$
 \\
 where $\mathbf{\lambda}^{-1/2}$ denotes the element-wise root inverse of the eigenvalues.\\
4. Calculate $\mathbf{K} = \mathbf{\Sigma}_{11}^{-1/2} \mathbf{\Sigma}_{12} \mathbf{\Sigma}_{22}^{-1/2}$.\\
5. Compute the SVD of $\mathbf{K}$ to get the largest $k$ singular values $\sigma$. \\
6. Return  $ \L = - \sum_{i=1}^{k} \sigma_i$.
\label{algo:b}
\end{algorithm}
\subsection{Gradient Computation}
The important aspect of the loss function is that it must be differentiable regarding its input. Andrew et al. \cite{andrew2013} provide a proof which is vital for our work as they prove that CCA is differentiable with respect to its inputs. In our case, we do not need the differentiation regarding our category indicator matrix which constitutes the second data vector for CCA. Still, we need the derivative to update our network weights for generating the first data vector. According to Andrew et al., the derivative for CCA is given by:
\begin{equation} \notag
\frac{\partial \mathrm{corr}(\mathbf{X},\mathbf{Y})}{\partial \mathbf{X}} = \frac{1}{m-1} (2 \nabla_{11} \bar{\mathbf{X}} + \nabla_{12} \bar{\mathbf{Y}}).
\end{equation}
Where $\bar{\mathbf{X}}$ and $\bar{\mathbf{Y}}$ are the centered data vectors and.
$\nabla_{11}$ and $\nabla_{12}$ are defined with respect to the singular value decomposition of $\mathbf{K} = \bm{\bm{\alpha}} \mathrm{diag}(\bm{\sigma}) \bm{\bm{\beta}}$ as in \cite{andrew2013}:
\begin{equation} \notag
\begin{split}
& \nabla_{12} = \mathbf{\Sigma}_{11}^{-1/2} \bm{\bm{\alpha}} \bm{\bm{\beta}}^T \mathbf{\Sigma}_{22}^{-1/2}, \\
& \nabla_{11} = -\frac{1}{2} \mathbf{\Sigma}_{11}^{-1/2} \bm{\bm{\alpha}} \mathrm{diag}(\bm{\sigma}) \bm{\bm{\alpha}}^T \mathbf{\Sigma}_{11}^{-1/2}. \\
\end{split}
\end{equation}

%We only require the derivative for the first view as we only update the network weights for the first view and not for both.
\subsection{Baseline Architecture} \label{sec:baselinearch}
So far, we have elaborated how CCA can be applied in our case in a CNN architecture in Section \ref{sec:DeepCCALDA}. Our baseline architecture extracts the first data vector in our proposed approach. The common choice in the deep hashing community is AlexNet \cite{alexnet2012} or CNN-F \cite{cnnf2014, hashnet2017, ddsh2018}. The convolutional layers of the CNN-F architecture are presented in Table \ref{tab:2}.  Local Response Normalization is applied on conv1 and conv2 layers. The network uses rectified linear unit as non-linearity and the convolutional layers are followed by two fully connected layers of dimension $4096$. The final fully connected layer consists of $\mathrm{C}$ nodes, which is the number of classes. AlexNet and CNN-F are usually pretrained on ImageNet such that they can be used as a feature extractor for images which depict a variety of concepts. Then the network is used for generating the first data vector $\mathbf{X}$. The second data vector $\mathbf{Y}$ is formed by the category indicator matrix and the network is trained with the derived loss function, as discussed in Section \ref{sec:loss}, to learn the low-dimensional feature space.
\begin{table}
\centering
\caption{\scriptsize The convolutional layers of the CNN-F architecture used as baseline architecture for generating the first data vector $\mathbf{X}$.}
\begin{tabular}{l c c c c c c c c}
    \toprule
    Layers  & conv1 & conv2 &conv3 &conv4 & conv5 \\
    Filters & 64 & 256& 256&  256 &  256 \\
    Fiter kernel &  11 $\times$ 11  &  5 $\times$ 5 &  3 $\times$3   &   3 $\times$3 &   3 $\times$3  \\ 
   % Dimension & 7744 & - &  - & - & - & 4096 &  4096& 1000\\  
    Stride  & 4 & 1 & 1 & 1 & 1 \\
    Padding & 0  &  2 & 1 & 1 & 1 & \\
    Max Pooling & 2 $\times$ 2  &  2 $\times$ 2  & - & - & 2 $\times$ 2  \\
    %Non-linearity & ReLU & ReLU& ReLU& ReLU & ReLU & ReLU& ReLU & -\\
    %ITQ \cite{itq2013} + CNN & 0.237 & 0.246 & 0.255 & 0.261 \\
    \bottomrule
\end{tabular}
\label{tab:2}
\end{table}
\section{Generation of Binary Codes} \label{sec:combinationwithitq} % 3 pages
Once we obtain an optimized low-dimensional feature space as discussed in Section 4, the next step is to binarize the feature vectors. We used the Iterative Quantization (ITQ) approach proposed by Gong et al. \cite{itq2013} for binarization of our feature vectors and briefly explain this approach here. In particular, we propose a combination of our devised feature extraction framework, Deep Canonical Correlation Analysis Feature (DCCF) extraction with ITQ. A crucial part of ITQ is the dimensionality reduction. Typically, PCA is used to project to a lower-dimensional space where a hypercube is fitted onto the feature vectors. However, CCA can also be used which was already proposed in the initial ITQ paper. The important criterion is that the projection is onto an orthogonal basis which is fulfilled by CCA as the projected features are uncorrelated. We use CCA loss to train our CNN. Thus, we have a CCA projection onto an orthogonal basis in our feature extraction framework. Hence, the criteria for the ITQ scheme are fulfilled. The central idea made in this paper is that instead of applying ITQ after a linear CCA, it can be naturally extended to our neural-network-based feature extraction method. The proposed DCCF method can therefore be considered as the dimensionality reduction step before ITQ. In short our DCCF extraction method provides:
\begin{enumerate}
\item An optimized feature space suitable for image retrieval.
\item A dimensionality reduction suitable for ITQ.
\end{enumerate}
\subsection{Iterative Quantization}
\label{sec:itq}
Here we explain the ITQ scheme briefly \cite{itq2013}.
We start with $q$ zero-centered feature vectors\footnote{As we will use ITQ in the context of image retrieval, we consider here directly the feature vectors $ \{ \mathbf{g}_1, \mathbf{g}_2, \dots, \mathbf{g}_q\}, \mathbf{g}_r \in \mathbb{R}^i$ from the gallery dataset.} $\mathbf{{g}_r} \in \mathbb{R}^{i}$.
These feature vectors form the rows of the data matrix $\mathbf{G} \in \mathbb{R}^{q \times i}$, after CCA projection. Here $q$ is the batchsize and $i$ is the length of the feature vector which corresponds to the number of correlation coefficients. Starting from this data matrix, we want to generate a binary code matrix $\mathbf{B} \in \{-1, +1\}^{q \times c}$, where $c$ is the length of the generated code.
As in the data-independent case, the $p^{th}$ bit belonging to the $r^{th}$ feature vector can be expressed as:
\begin{equation} \label{eq:itqsingleb}
b_{r,p} = h_p(\mathbf{g}_r) = \mathrm{sign}(\mathbf{g}_r \mathbf{w}_p),
\end{equation}
where $h$ denotes the hashing function.
The vectors $\mathbf{w}_p$ denote the projections we want to learn. 
The sign-function on a matrix is the element-wise sign-function, and we can express \eqref{eq:itqsingleb} for all bits, i.e., the entire binary code, in the matrix form as:
\begin{equation} \label{binarycodes}
\mathbf{B} = \mathrm{sign}(\mathbf{G} \mathbf{W} \mathbf{R} ) \text{, where } \mathbf{W} \in \mathbb{R}^{i \times c},
\end{equation}
where $\mathbf{W}$ denotes the matrix which contains the hashing hyperplanes. $\mathbf{R}$ is the optimal rotation matrix which rotates the data such that the quantization loss, $Q$ of the feature vectors is minimized as given below:
\begin{equation} \label{eq:itqminfrobenius}
Q(\mathbf{B},\mathbf{R}) = \norm{ \mathbf{B} - \mathbf{GW}\mathbf{R}}_F^2 \cdot
\end{equation}
Here, $\norm{ \cdot }_F$ denotes the Frobenius norm of the matrix. Gong et al. initialize the rotation matrix $\mathbf{R}$ with a random matrix since it is proven that random projection distributes the variance effectively \cite{jegou2010}.\par
In this iterative scheme, first, we will look for an optimal $\mathbf{B}$ given the random initialization matrix $\mathbf{R}$ from \eqref{binarycodes}. Once we have updated $\mathbf{B}$, the $\mathbf{R}$ matrix is updated such that \eqref{eq:itqminfrobenius} is minimized. Finding the optimal $\mathbf{R}$ according to \eqref{eq:itqminfrobenius} becomes easy when we remember the orthogonal Procrustes problem. In the orthogonal Procrustes problem, we are trying to find an optimal rotation to align two sets of points. Here, for a fixed $\mathbf{B}$, \eqref{eq:itqminfrobenius} is exactly this orthogonal Procrustes problem as we are looking for a rotation matrix that aligns the matrices $\mathbf{B}$ and $\mathbf{GWR}$. Consequently, the solution can be found by an SVD which provides the solution to the orthogonal Procrustes problem as $\mathbf{B}^T \mathbf{GW} = \mathbf{U} \mathbf{\Sigma} \mathbf{V}^T$ and then we can directly update $\mathbf{R}$ as $\mathbf{V} \mathbf{U}^T$. These steps are repeated several times iteratively to find the final binary code $\mathbf{B}$.
\subsection{Combination with ITQ}
Once we find an optimized feature space with our Deep Canonical Correlation Feature (DCCF) extraction, we can combine it with an unsupervised binarization scheme to generate effective and efficient binary codes for image retrieval. The main advantage of this two-step approach is that while we need labels for the supervised feature learning, the quantization loss can be minimized on the unlabeled feature vectors as well. Consequently, we need a small labeled subset of the gallery dataset to train our CNN by our proposed loss function. For the binarization step, we can use all gallery images.
We can directly find low-dimensional feature representations of the images by our proposed method, instead of first generating a high-dimensional feature space followed by a dimensionality reduction.
The dimensionality of our final features are bounded by the number of classes and by the fact whether they are mutually exclusive. The dimensionality of our final feature vector is:
\begin{equation} \notag
i \leq
\begin{cases}
\mathrm{C-1}, \text{ if the classes are mutually exclusive}\\
\mathrm{C}, \text{ otherwise}.\\
\end{cases}
\end{equation}\par
\begin{figure}
\begin{center}
% scale = 0.3
  \includegraphics[height = 0.21\textwidth , width = 0.49\textwidth]{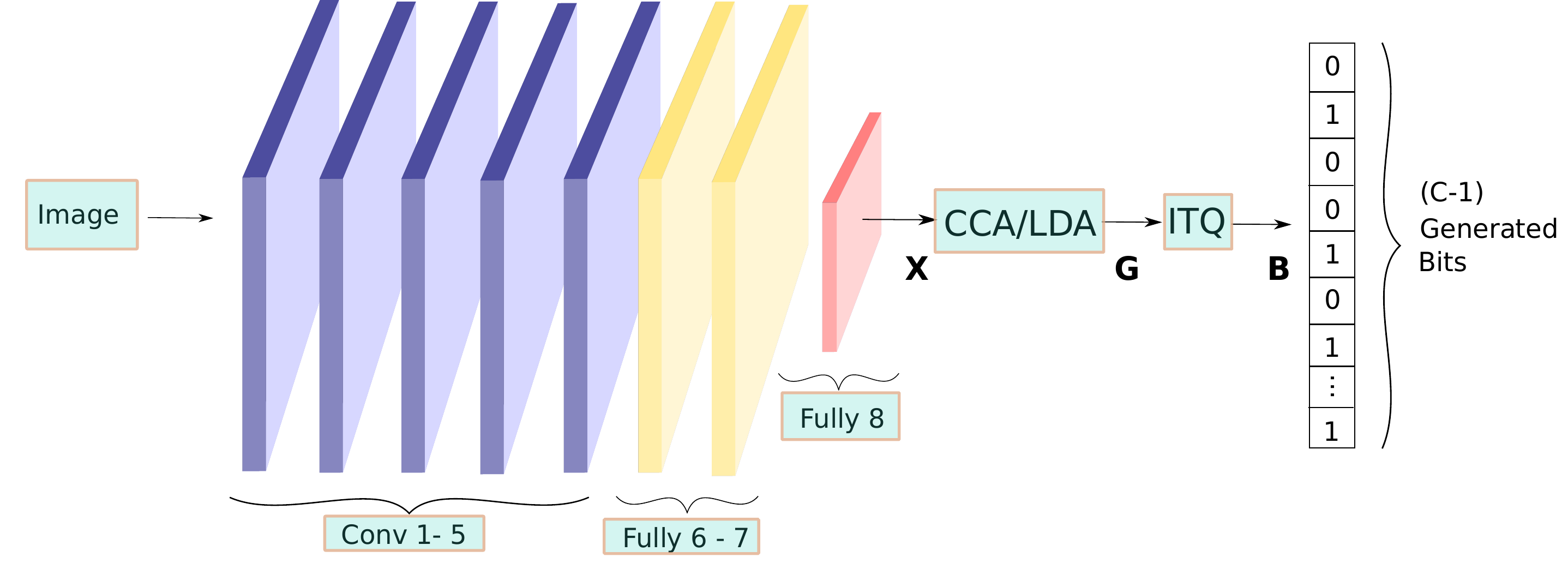}
  \captionsetup{justification=centering}
  \caption{\scriptsize Our proposed feature extraction method followed by the binarization with ITQ. The feature vectors $\mathbf{X}$ are extracted from the fully connected layer and after applying CCA, we get the projected feaure vectors $\mathbf{G}$. These features are then binarized by ITQ to obtain the binarized codes $\mathbf{B}$. The full binarization pipeline is reffered in this paper as Deep Canonical Correlation Hashing (DCCH).}
  \label{fig:dcliMETHOD}
\end{center}
\end{figure}

Note that this is only an upper bound for the dimensionality. Of course, we can always project to a low-dimensional space.
In the following section, we assume that our feature vectors are in an $i$-dimensional space.
By applying ITQ on the gallery images, we try to fit an $i$-dimensional hypercube to our feature vectors. We can find $i$-bits with a low-correlation since the hashing hyperplanes are orthogonal \cite{itq2013}. In our case they are orthogonal since the rotation matrix, $\mathbf{R}$ found by ITQ is orthogonal as well as our projected feature vectors generated by the network. Our final binarization method with ITQ is depicted in Fig. \ref{fig:dcliMETHOD}. For the images, we compute the feature embedding by our DCCF method. Then, we rotate the feature vector matrix according to the learned projection matrix by ITQ. Finally, we use the sign thresholding by ITQ to generate the binary codes for the given input image. The combination of our DCCF method with ITQ is referred in this paper as Deep Canonical Correlation Hashing (DCCH).
\section{Network Ensemble} \label{sec:networkensemble}
The combination with ITQ leads to generation of efficient binary codes for image retrieval. However, as we use our DCCF method, we are bound by the number of mutually exclusive classes. For instance, in the case of CIFAR-10, we could only project to a 9-dimensional feature space. Thus, we could generate only $9$ bits by ITQ. Since, the benchmarks of the current state-of-the-art methods use higher number of bits such as $12$, $16$, $24$, $32$, $48$ and $64$ \cite{ddsh2018}, we need to generate more bits for a fair comparison.
%Gong et al proposed a double bit quantization scheme applicable for ITQ. There, they spend more than two bits per dimension to better respect the neighboring structure. However, if look at the toy data example from Section again as depicted in Fig. NUMBER.
%\begin{figure}
%\centering
%\includegraphics[scale=0.5]{fig/fundamentals/beforeITQFig.pdf}
%\caption{Toy Data Points for a Binarization}
%\label{fig:DeepCCAToyITQ}
%\end{figure}
%There, we can clearly see that for some red points some green points are closer than all other red points. Having a more accurate euclidean-distance preserving double-bit quantization does not make sense consequently. This can be clearly seen in Fig. NUMBER as well when we only consider one of the projected dimensions.
We propose an ensemble technique to increase the bit size. The idea is that during the training, randomness is involved. The sampled covariance matrices which are needed in the loss calculation of our DCCF method are constructed based on our random batches. Since the images within a batch are selected randomly, the estimation of the covariance matrices differs during each iteration. Consequently, the learned feature vectors differ. The binarization through ITQ also involves randomness due to the initial random projection $\mathbf{R}$ which is used for distributing the variance across the different feature dimensions effectively (cf. Sec. \ref{sec:itq}). Therefore, the bits generated through different trainings on the same training data and the binarization will result in different binary codes during each iteration.\par The same observation is commonly utilized for classification~\cite{networkensembleclassification}. The central idea is that different networks vote for the classes and the major vote will be the final classification result. For image retrieval, ensemble techniques were proposed by Huang et al. \cite{ensembleImageRetrievalAverageFeatures2016}. These ensemble techniques mainly aim at finding features by an ensemble of multiple networks with a different structure. Our proposal is different as we use the ensemble of the same baseline architecture to generate the final sub codes which are simply concatenated afterwards. By doing this, our ensemble technique can be interpreted as a voting among networks.\par
\begin{figure} [h]
\begin{center}
  \includegraphics[height = 0.22\textwidth , width = 0.49\textwidth]{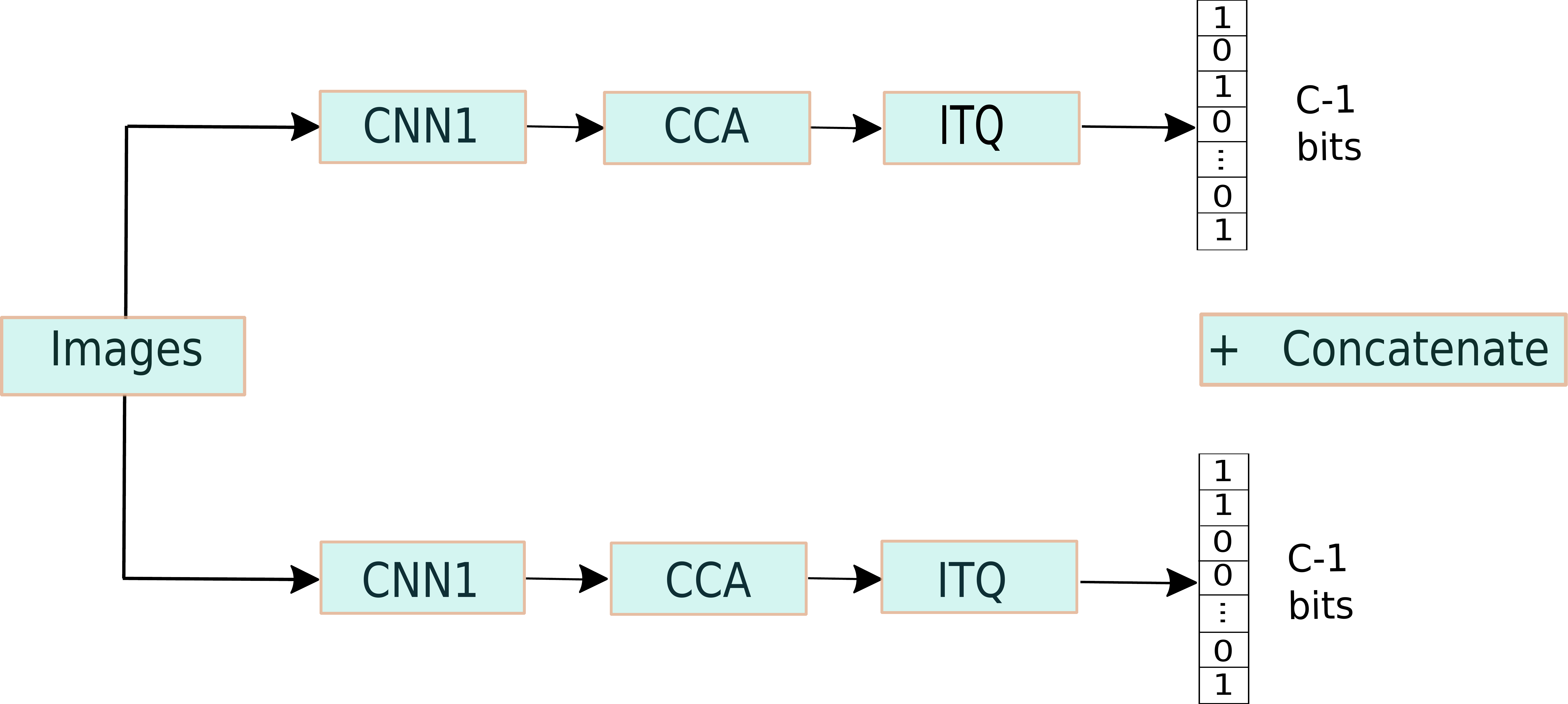}
  \captionsetup{justification=centering}
  \caption{\scriptsize Ensemble network to increase the bitlength by concatenating the bits from different networks.}
  \label{fig:ensembleMETHOD}
\end{center}
\end{figure}

 If we want to generate more than $\mathrm{C-1}$ bits with our DCCH method, we need to use an ensemble technique. However, if we definitely know that $\mathrm{C-1}$ is greater than the desired number of bits, no ensemble technique is required.
Our ensemble approach for an ensemble of two networks is depicted in Fig. \ref{fig:ensembleMETHOD}. We know by the properties of ITQ that the bits generated by each network are uncorrelated to other bits generated by the same network. However, if we concatenate the bits of the different networks, we cannot ensure that the bits between the networks are uncorrelated. Therefore, we propose to adapt the dimensionality reduction for binary features by Rublee et al. to our ensemble technique \cite{orb2011}.
Rublee et al. discuss the case where we have many binary tests and we want to choose only a few of them which is a dimensionality reduction in the form of $\mathbb{B}^i \mapsto \mathbb{B}^n$. The selected bits, should be uncorrelated to create an effective subset of the original bits. The authors propose a greedy search scheme which we can adapt to our method as explained in Algorithm \ref{algo:2} below:
\begin{algorithm}
\caption{. Greedy search scheme for generating longer codes}
%\DontPrintSemicolon % Some LaTeX compilers require you to use \dontprintsemicolon    instead

1. Choose the $\mathrm{i}$ bits from any network from the ensemble and compute the matrix, $\mathbf{M} \in \mathbb{R}^{q \times i}$, where $q$ is the batch size.\\
2. Choose the next bit from any other network and form a vector, $v \in \mathbb{R}^{1 \times i}$.\\
3. Measure the correlation of $v$ with each row of $\mathbf{M}$.\\
4. If the correlation is below a certain threshold concatenate it to your bitcode.\\ 
5. If we have found the desired number of bits, then stop.\\ 
6. Else, continue with step $2$.
\label{algo:2}
\end{algorithm}
\par
If the search found fewer bits than the desired number of bits, the threshold for the correlation among the bits has to be decreased. Note that the procedure for selecting the bits only has to be done once on the gallery dataset. For a query image, we can then generate the bits from the outputs of the network corresponding to the bits which were chosen by the greedy search. By doing this, we reduce the correlation among the bits generated from our ensemble network. In the case of CIFAR-10, we could, e.g., train five networks for the feature space generation and the following ITQ. Then we choose the bits of one of these networks. We could then select bits from the other networks which are uncorrelated with our already chosen bits according to the greedy search scheme. Note that it makes sense that if our $10^{th}$ bit is chosen from a network, we should first look for other bits from this particular network as these bits are uncorrelated to the $10^{th}$ bit. The greedy search scheme can also be performed on the data of all gallery images because no labels are required for estimating the correlation among the bits.
\section{Experiments}
Deep Discrete Supervised Hashing (DDSH)\footnote{To the best of our knowledge, DDSH also presents the best results achieved so far under the common image retrieval benchmarks, so that we compare our method to one of the best existing deep hashing methods.} \cite{ddsh2018} is one of the most recent state-of-the-art hashing methods and we follow exactly their experimental setup for the datasets CIFAR-10 (single-labeled) and NUS-WIDE (multi-labeled). To have a fair comparison, we use the same baseline architecture, which is CNN-F as the feature extractor. We also use the same data splits for the training, query, and gallery datasets as described by the authors of DDSH. All benchmarks involving a GPU were conducted on a GeForce GTX TITAN-X with 12 GB memory. We have used the parameters given in Table \ref{tab:parameters} for our experiments which were found by a random grid search \cite{grid}.  The conducted benchmarks are especially interesting in the case of multi-labeled images since we addressed the extension of DeepLDA to multi-labeled images by our method. Therefore, we conduct experiments in another multi-labeled dataset, the MS-COCO dataset \cite{mscoco2014}. DDSH does not provide the results for MS-COCO. So, we use the commonly used HashNet\cite{hashnet2017} as the baseline for the comparison in this case.
\subsection{Evaluations}
In this section, we summarize the experimental evaluations. We have plotted the training loss for all the experiments. We have also visualized the feature space to see the distribution of feature embeddings learned by our feature extraction technique. In all the three datasets, we have measured the mean average precision to measure the retrieval performance.
\subsubsection{Training Loss and Visualization of Feature Space}
One crucial aspect of our method is its feature extraction. For learning a good feature space, we train our model with the training objective as given in \eqref{eq:ourloss}.
% \begin{equation} \label{eq:ResEq1}
% Loss = -\sum_{i=1}^{k} \rho_i = -\sum_{i=1}^{k} \sqrt{ \frac{\lambda_{LDA,i}}{1+ \lambda_{LDA,i}} }
% \end{equation}
%where $\rho$ is the correlation coefficient and $\lambda_{LDA}$ the eigenvalue in case of the LDA equivalence. This leads to a helpful intuition behind our loss since the correlation coefficient in each dimension is bounded by $1$ as discussed in Section \ref{sec:ccabysvd}. Therefore, we can interpret the training loss to see if our method was able to find a good feature representation. With the bounds of the correlation coefficient and from Eq. \ref{eq:ResEq1}, we can directly infer a minimum loss value which provides a lower bound for the training loss, which is:
%\begin{equation} \label{eq:ResEq2}
%Loss_{min} = -k
%\end{equation}
When we get close to this minimum bound, we could infer that we trained our model in the optimum way. Thus, we will inspect the training loss for the mentioned datasets to evaluate our results. Once our model is trained till the loss reaches the lower bound, a feature space suitable for image retrieval is generated. We use t-Distributed Stochastic Neighbor Embedding (t-SNE) \cite{tsne2008}, a popular tool to visualize the feature space of the training images in 2D. 
%This method is a nonlinear dimensionality reduction method where we can project data points which are close in the original feature space to nearby points in the resulting space with a high probability. For instance, we can use our DCL method to generate a 9-dimensional feature space for CIFAR-10.
% To visualize this feature embedding, we can use t-SNE to project to a 2-dimensional space which preserves the neighboring structure. 
\subsubsection{Mean Average Precision}
Our final objective is the generation of binary codes suitable for image retrieval. Thus, we need benchmarks which can assess the performance of image retrieval frameworks. The essential characteristics of an image retrieval system for a given query image are its precision and recall. We measure these characteristics with respect to the Hamming ranking of the gallery images for a given query image \cite{smeulders2000}. For single-labeled images, the relevance of an image indicates if the retrieved image belongs to the same class \cite{cbirsurvey2008}. For multi-labeled images, a retrieved image is usually defined as relevant if it shares at least one category label with the query image \cite{ddsh2018}. Precision is defined as the ratio of the number of relevant retrieved images to the total number of retrieved images. Besides the precision of our retrieved results, recall is the second important measure \cite{smeulders2000}. Recall is the ratio of the relevant retrieved images to all relevant images in our gallery dataset \cite{cbirsurvey2008}.
The mAP-score is the mean of the Average Precision (AP). AP score is defined as the average of precision when recall varies from 0 to 1. In general, we will report the average of applying our experiment five times for a more stable mAP-score.
\subsection{Experimental Results on the CIFAR-10 Dataset}
The CIFAR-10 \cite{cifar10} dataset is widely used for benchmarking the performance of machine learning algorithms. It comprises $10$ classes which are mutually exclusive. CIFAR-10 comprises $60{,}000$ images which are of size $32\times32$. The typical training/test split consists of $50{,}000$ training images and $10{,}000$ test images.\par However, using such a large training dataset in comparison to a small test dataset is unreasonable since the retrieval method should only fine-tune an existing feature extractor on a small training subset. If all training images are used for the training and later constitute the gallery dataset, image retrieval could be substituted by a simple classification \cite{howshouldwe2017}. Consequently, the current image retrieval benchmarks use a training set of only $5{,}000$ training images for CIFAR-10 \cite{ddsh2018, dtsh2016, hashnet2017}. Following DDSH \cite{ddsh2018}, $500$ images are randomly sampled from the dataset for each class label resulting in $5{,}000$ training images. Furthermore, $100$ images per class are randomly sampled for the query dataset which comprises $1{,}000$ images in total. We adopt the usual approach that the gallery dataset includes the training dataset as in the case of DDSH \cite{ddsh2018}. CIFAR-10 contains single-labeled images. Thus, the equivalence to LDA holds in this case. So, we can project maximally to a 9-dimensional space. Hence, we use the ensemble technique proposed in Section \ref{sec:networkensemble} for generating longer binary codes.
\begin{figure}
    \begin{subfigure}{0.25\textwidth}
            \includegraphics[height = 0.67\textwidth , width = 1\textwidth  ]{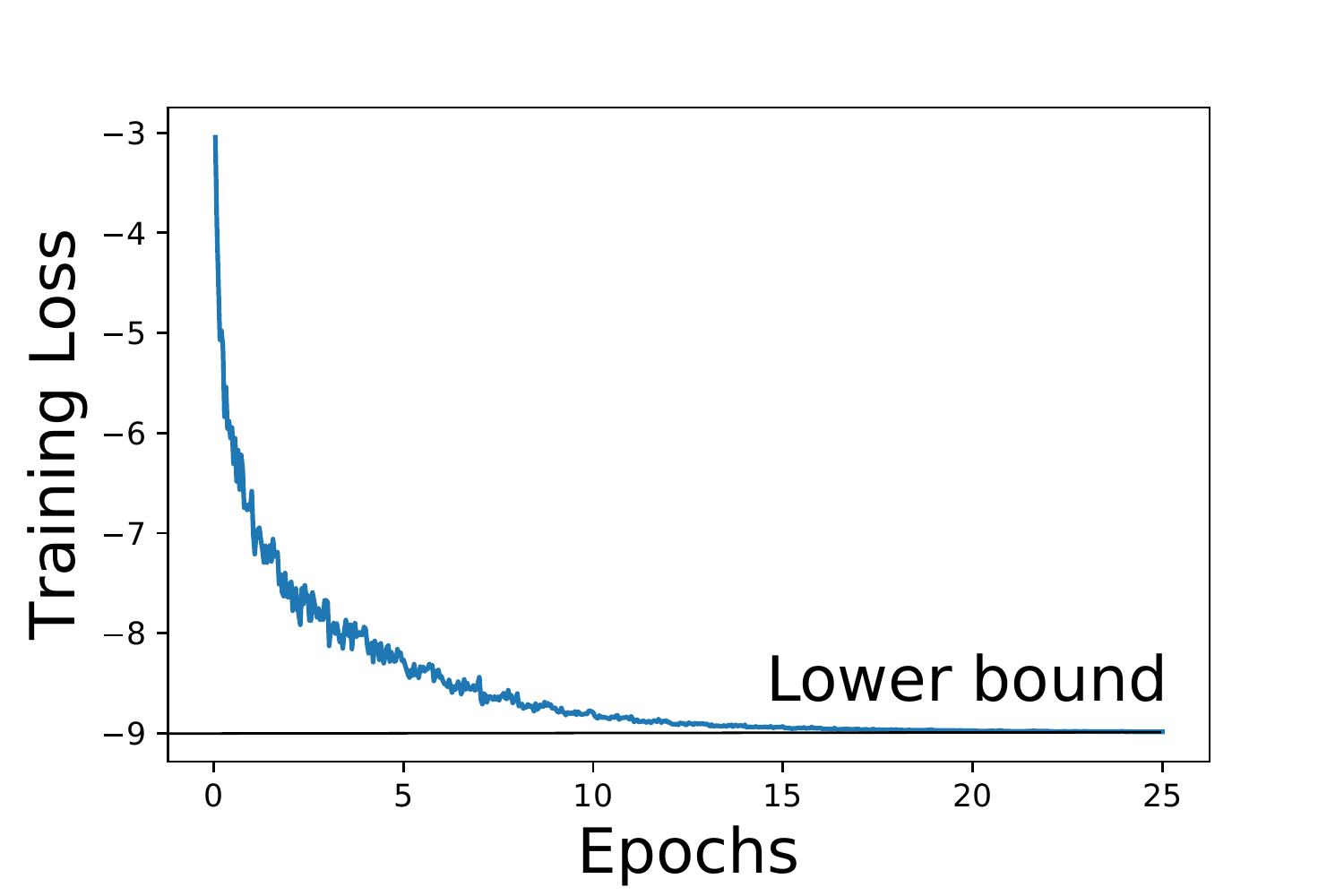}
             \subcaption{ \scriptsize CIFAR-10}
             \label{fig:closs}
             \includegraphics[height = 0.67\textwidth , width = 1\textwidth , scale=.25]{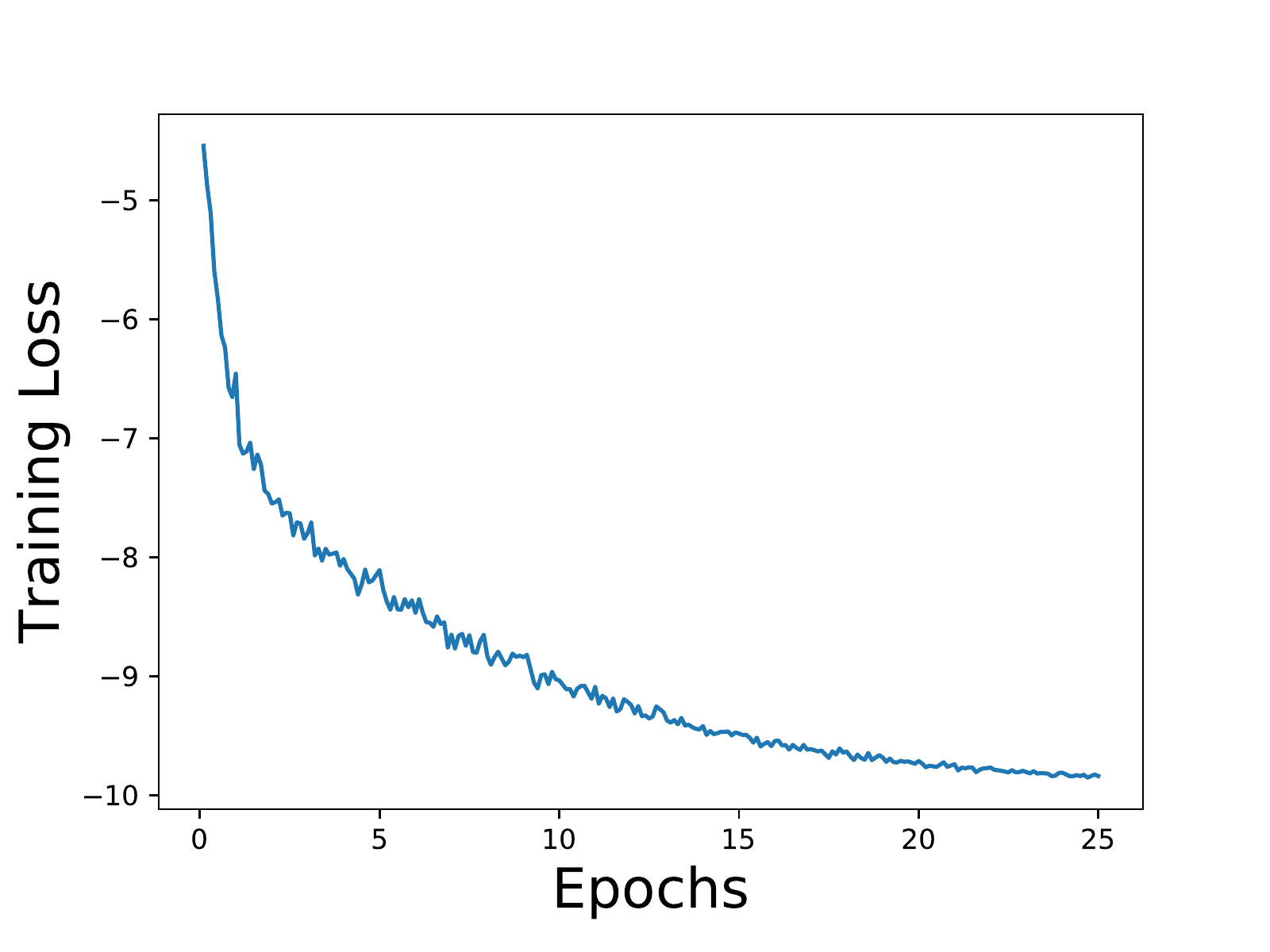}
             \subcaption{\scriptsize  NUS-WIDE}  
              \label{fig:nusloss}
    \end{subfigure}%
     %add desired spacing between images, e. g. ~, \quad, \qquad etc.
    \begin{subfigure}{0.25\textwidth}
             \includegraphics[scale=.3]{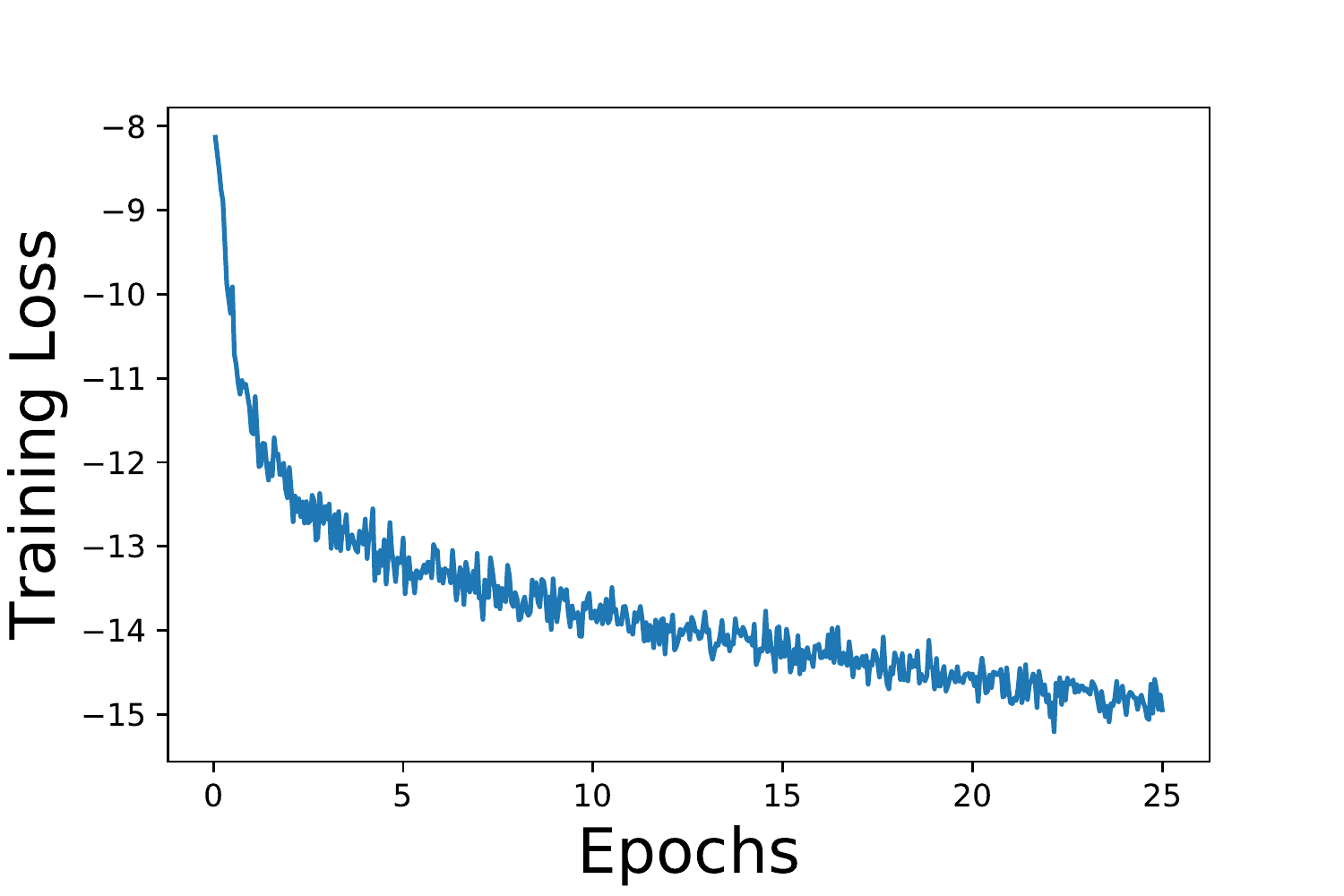}
             \scriptsize \subcaption{\scriptsize  MSCOCO 16 dimensional}
              \label{fig:mscoco17loss}
             \includegraphics[scale=.3]{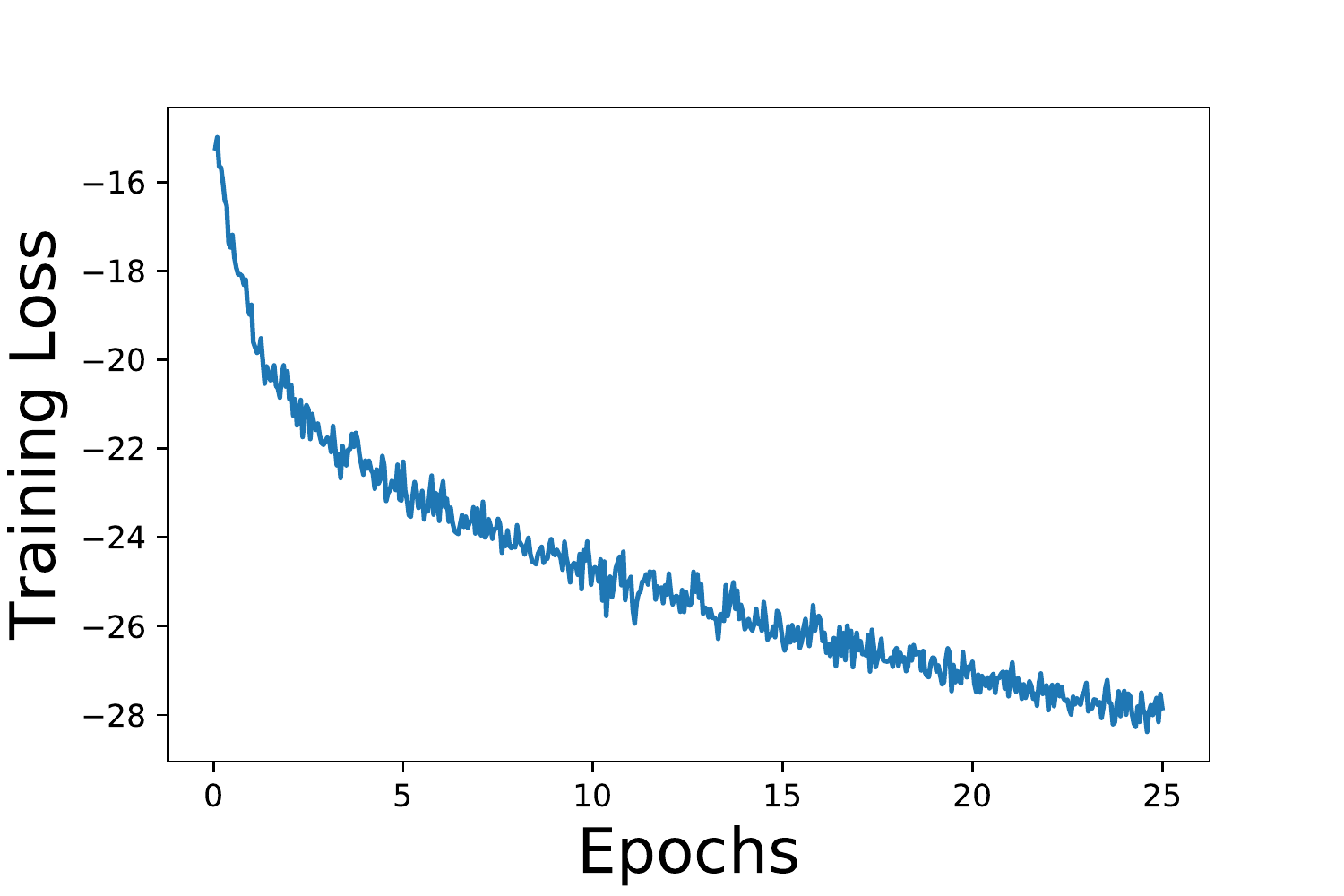}
              \scriptsize \subcaption{\scriptsize  MSCOCO 32 dimensional}  
               \label{fig:mscoco33loss}
    \end{subfigure}
    %\label{fig:loss}
    \caption{\scriptsize (a) Training loss for CIFAR-10. We could clearly observe that the loss goes to the lower bound of -$9$ which is the sum of the correlation coefficients $\rho$ in the $9$ directions which is possible since it is a single-labeled image dataset, (b) training loss on NUS-WIDE for 10 dimensional space, (c) training loss on MSCOCO dataset for 16 dimensional space, (d) training loss on MSCOCO dataset for 32 dimensional space.}\label{fig:loss}
\end{figure}
\begin{table}[h]
    \caption{\scriptsize Training parameters for CIFAR-10, NUS-WIDE and MS-COCO datasets.}
    \centering
    \begin{tabular}{l l l l l}
      \toprule
      Parameter & CIFAR-10  &  NUS-WIDE &  MS-COCO \\
      \midrule
      Feature Extractor &  CNN-F  &  CNN-F  &   AlexNet   \\
      Batch Size  &  200     &  500  & 500 \\
      Optimizer &  SGD  & SGD  & SGD \\
      Learning Rate &  1e-3   &  1e-3  &  1e-3  \\
      Epochs & 25  & 25 & 25 \\
      Correlation coe. & 9  & 10 & bits \\
      Ensemble & yes  &yes & no \\
            \bottomrule
       \label{tab:parameters}
    \end{tabular}
  \end{table}
\subsubsection{Training Loss and Learned Feature Space}
\begin{figure*}[h]\centering
  \begin{subfigure}[t]{0.33\textwidth}\centering
    \includegraphics[width=\textwidth,height=\textheight,keepaspectratio]{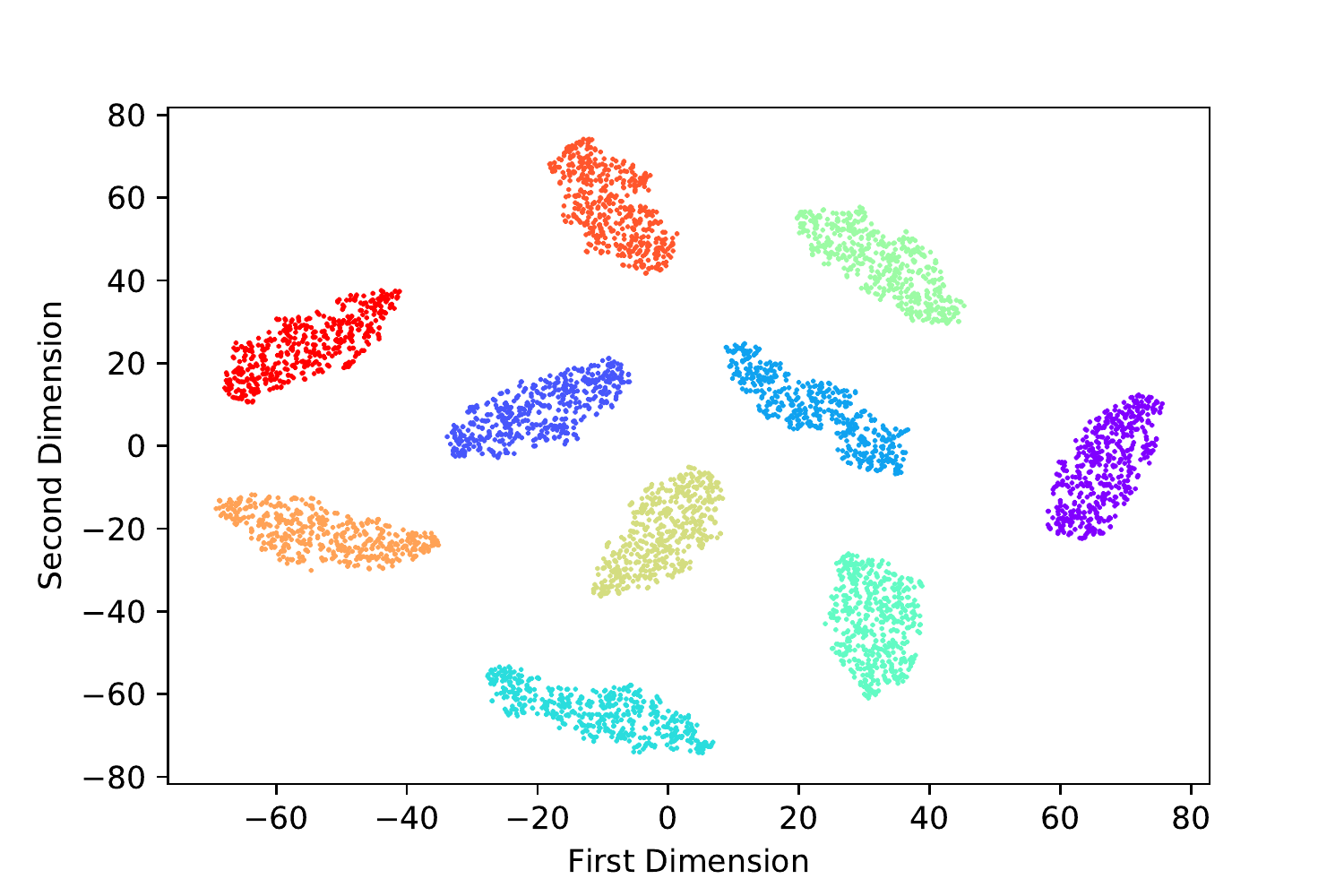}
    \caption{CIFAR-$10$ dataset}\label{fig:cifar-10}
  \end{subfigure}
       \begin{subfigure}[t]{0.33\textwidth}\centering
     \includegraphics[width=\textwidth,height=\textheight,keepaspectratio]{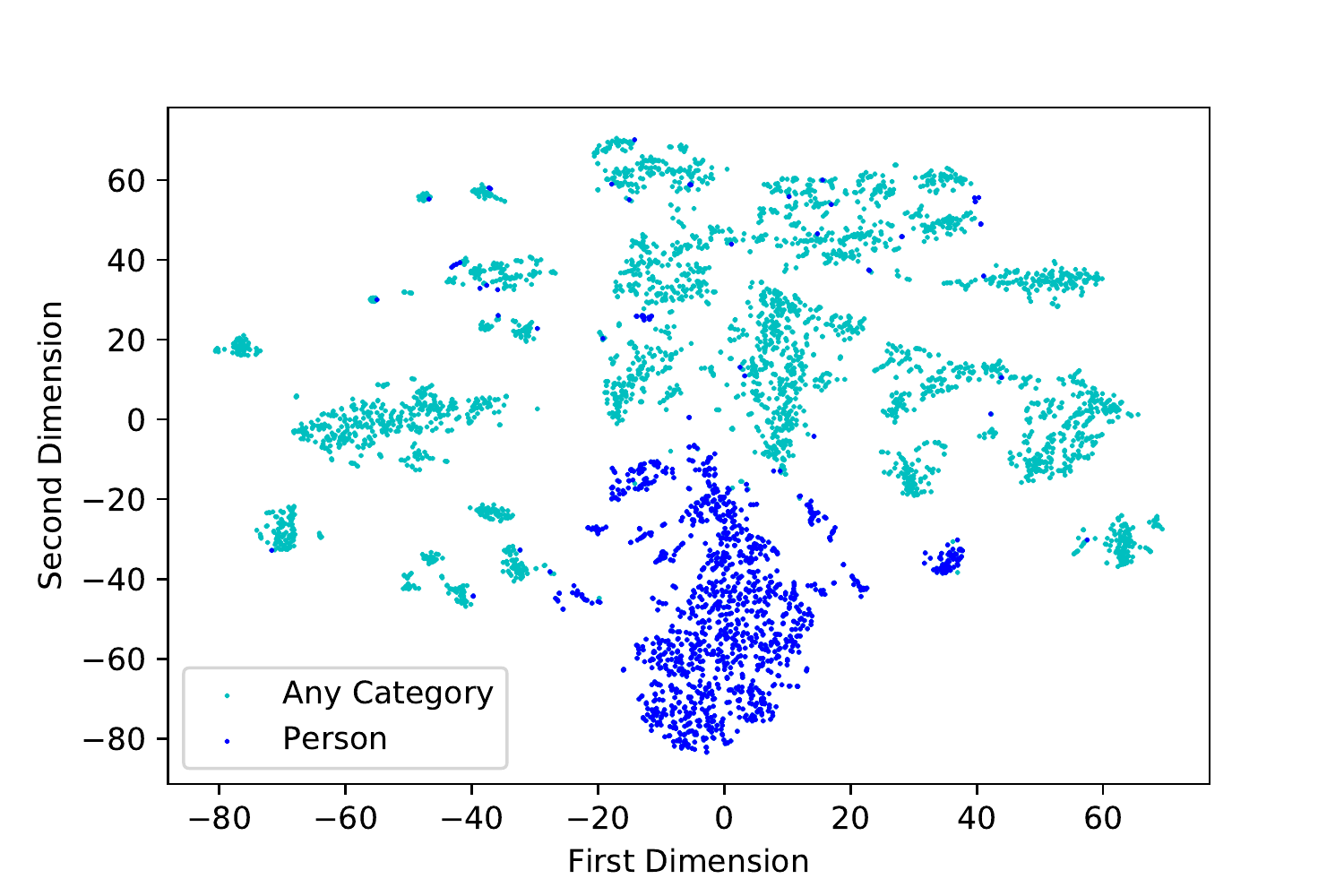}
      \caption{NUS-WIDE person category}\label{fig:person}
  \end{subfigure}
      \begin{subfigure}[t]{0.33\textwidth}\centering
    \includegraphics[width=\textwidth,height=\textheight,keepaspectratio]{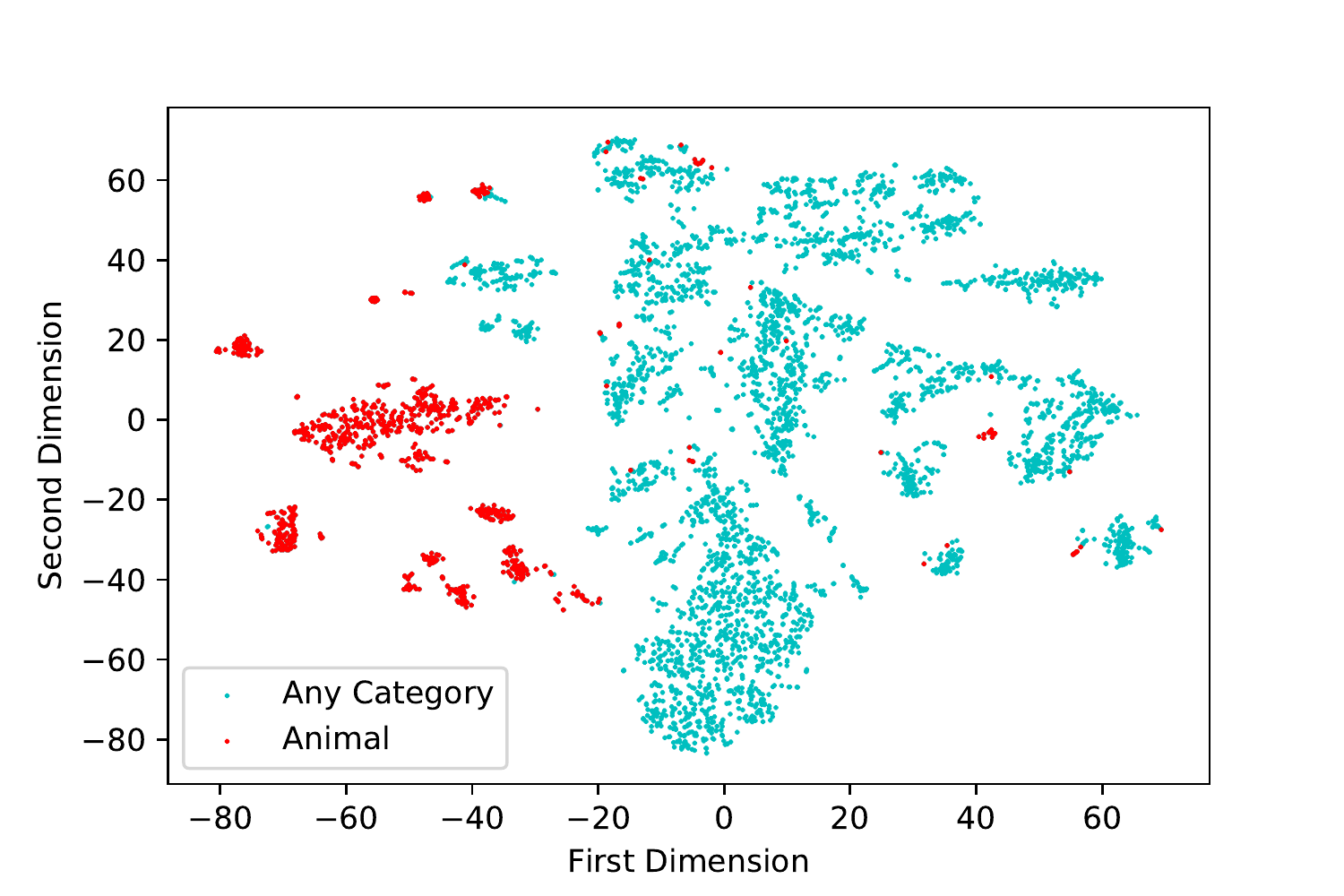}
    \caption{NUS-WIDE animal category}\label{fig:animal}
  \end{subfigure}
    \begin{subfigure}[t]{0.33\textwidth}\centering
	\includegraphics[width=\textwidth,height=\textheight,keepaspectratio]{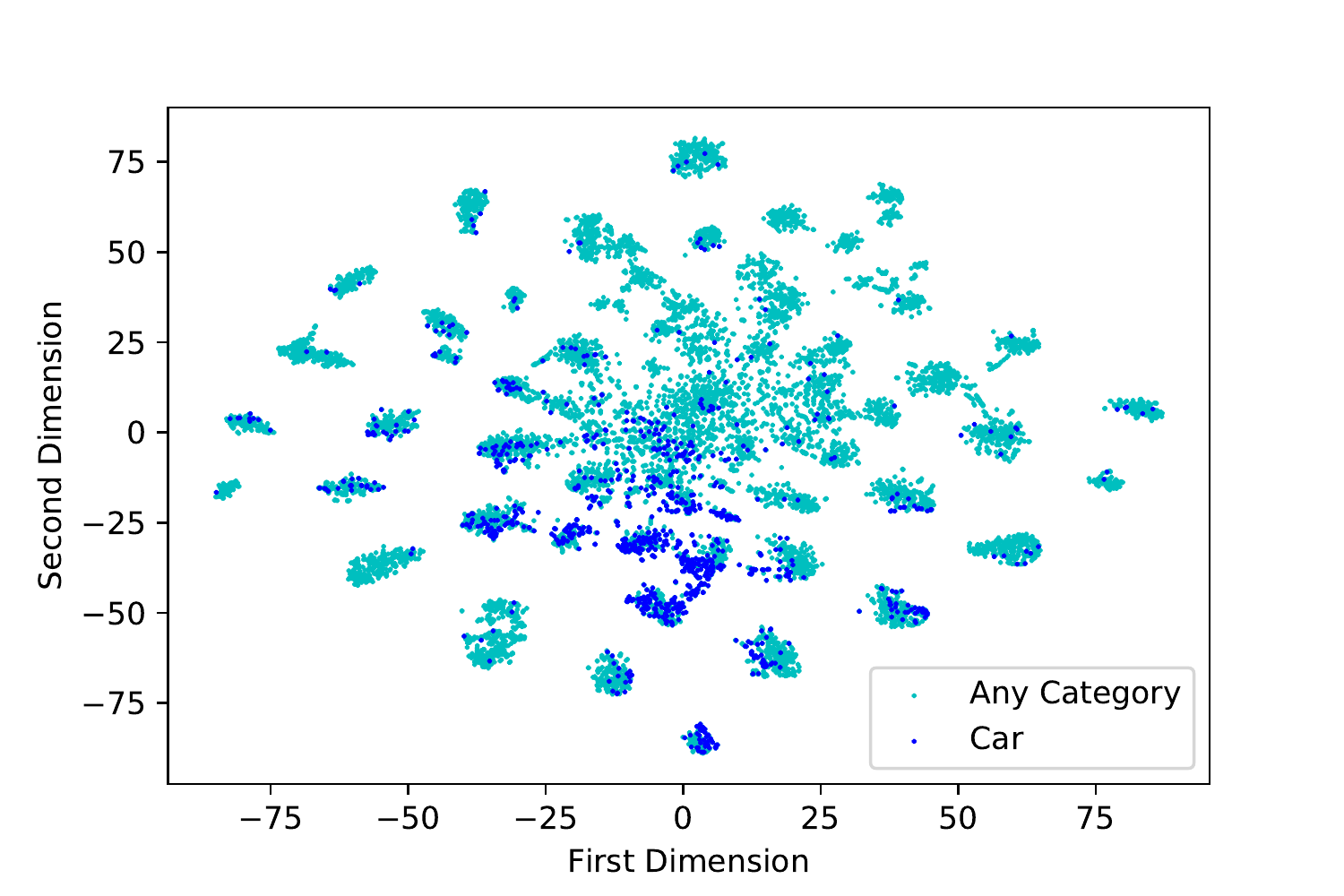}
	\caption{MSCOCO car category}\label{fig:car}
\end{subfigure}
      \begin{subfigure}[t]{0.33\textwidth}\centering
 \includegraphics[width=\textwidth,height=\textheight,keepaspectratio]{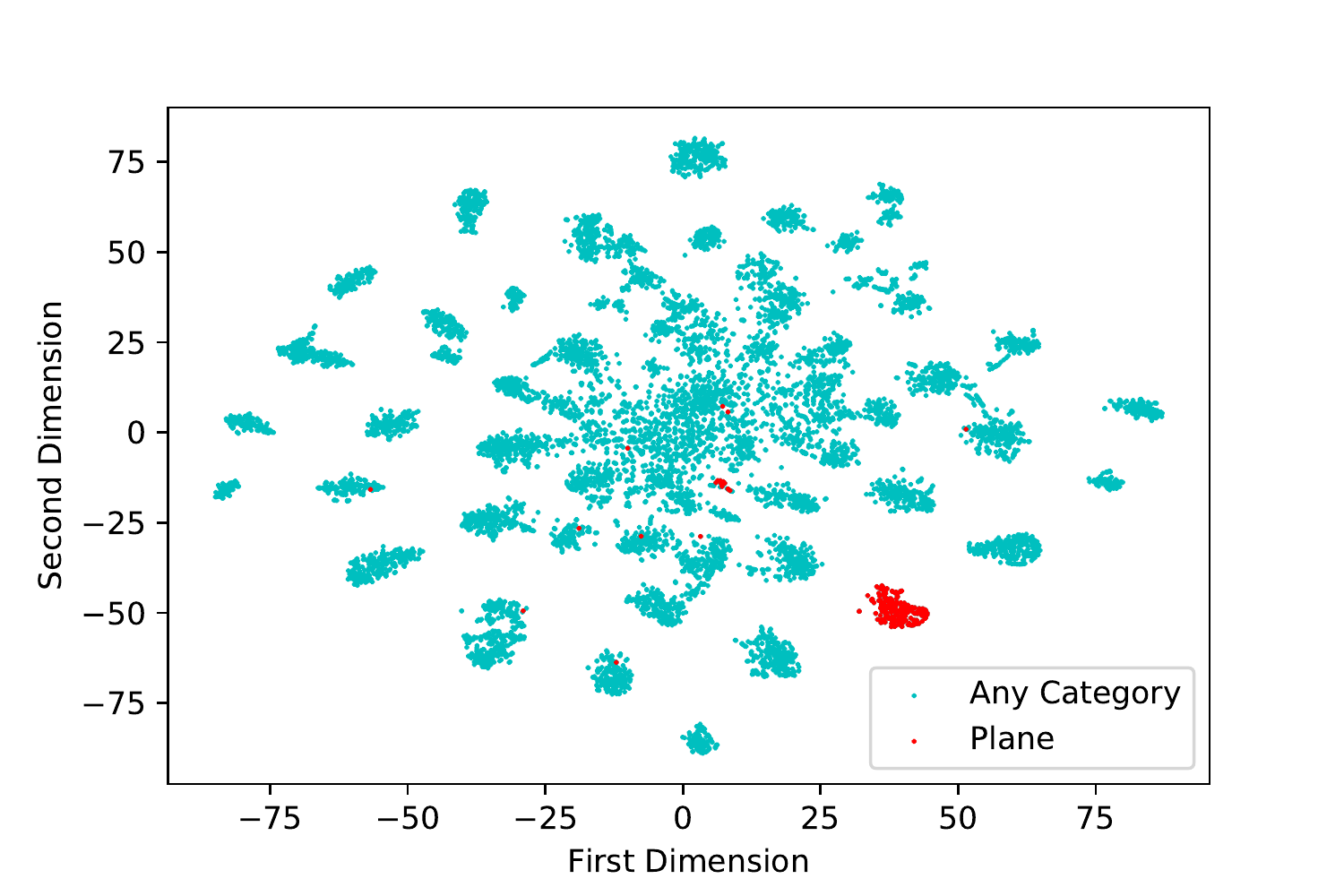}
    \caption{MSCOCO plane category}\label{fig:plane}
  \end{subfigure}
    \begin{subfigure}[t]{0.33\textwidth}\centering
    \includegraphics[width=\textwidth,height=\textheight,keepaspectratio]{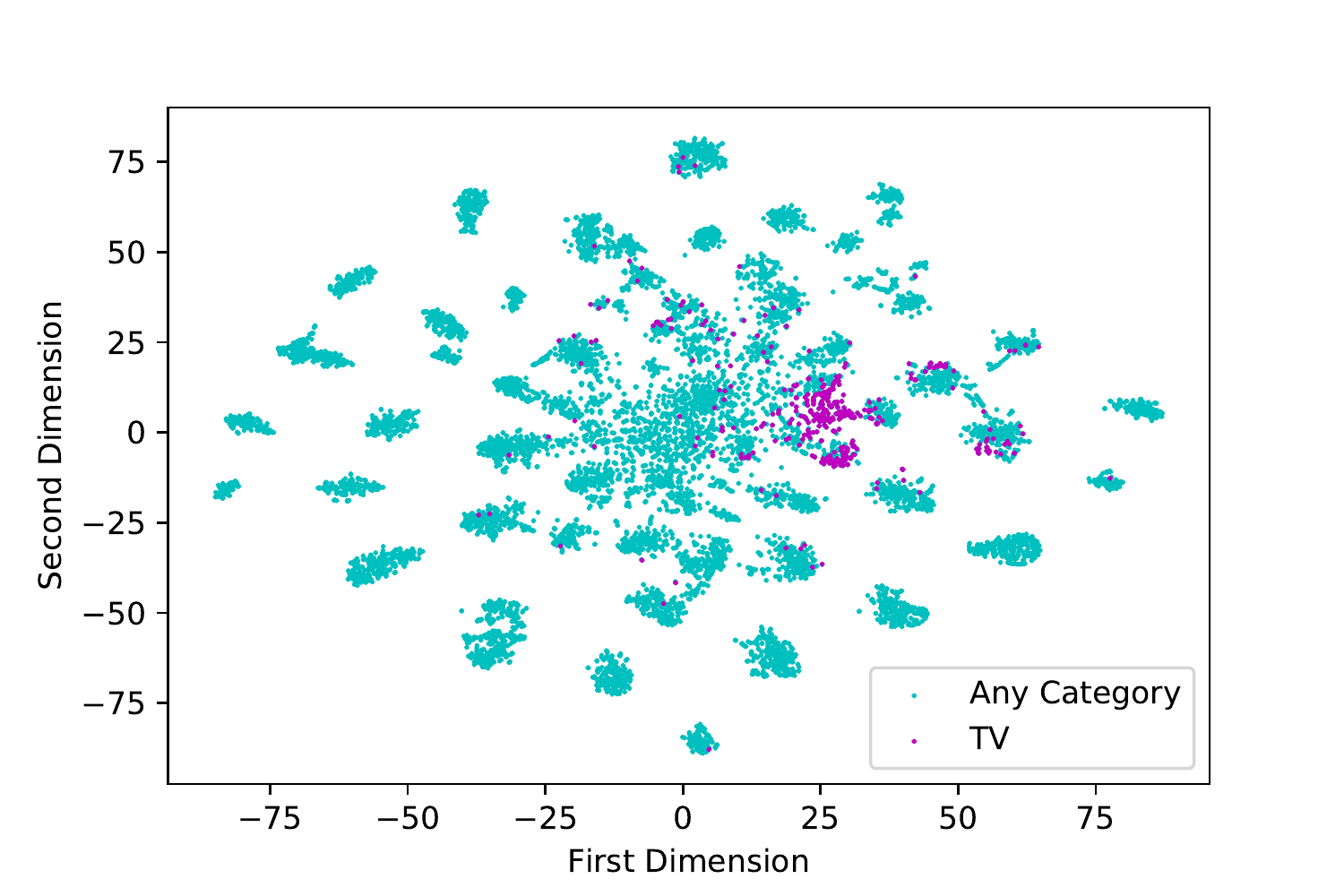}
    \caption{MSCOCO TV category}\label{fig:tv}
  \end{subfigure}
  \caption{\scriptsize Two dimensional visualization of feature space using t-SNE for, (a) CIFAR-10 dataset (Due to the equivalence between LDA and CCA, the classes are well separated), (b) NUS-WIDE highlighting person category (We could observe a big cluster formation for this category),  (c) NUS-WIDE highlighting animal category (Most of the image features belonging to this class are clustered together. Some feature points belonging to animal category are further away showing the presence of images which contain this category, but animal category is not the dominant category for those images) (d) MSCOCO highlighting plane category (Clusters all images which contains plane images together), (e) MSCOCO highlighting car category (Does not form a compact cluster indicating that this category is not dominat in most of the images), (f) MSCOCO highlighting TV category which are clustered together.}\label{fig:big_fig}
\end{figure*}

In Fig. \ref{fig:closs}, the training loss for our proposed method is depicted. For single-labeled images, the equivalence between LDA and CCA holds, and therefore the index of correlation coefficients is $\mathrm{C}-1 = 9$.  The maximum value for each correlation coefficient is $1$. We can clearly observe that we reach the lower bound of $\mathrm{loss_{min}} = -9$. As we use $9$ correlation coefficients by projecting to $9$ dimensions, this is the minimum value which can be achieved. It indicates that our training has converged properly substantiating the generation of an optimized feature space. In Fig. \ref{fig:cifar-10},
the 2-dimensional t-SNE visualization of our 9-dimensional feature space is illustrated. We depict the feature space for the training images based on category information. Different colors represent different classes. Note that the 2-dimensional embedding by t-SNE cannot capture the complete neighboring structure in the 9-dimensional space which we use for binarization. Still, we can see that a nearest neighbor search on such a feature space would lead to excellent retrieval results. Furthermore, this feature embedding shares the properties of DeepLDA due to the equivalence of CCA and LDA for single-labeled images, i.e., we have a high inter-class variance and a small intra-class variance.
\subsubsection{Mean Average Precision}
In Table \ref{tab:mapCIFAR10}, the mAP-scores are summarized for different bit lengths. Deep Discrete Supervised Hashing (DDSH) \cite{ddsh2018}, Deep Supervised Discrete Hashing (DSDH) \cite{dsdh2017}, Deep Supervised Hashing with Triplet Labels (DTSH)\cite{dtsh2016}, Deep Supervised Hashing with Pairwise Labels (DPSH)\cite{dpsh2015}, Data Sensitive
Hashing (DSH)\cite{dsh2016}, Deep Hashing Network (DHN)\cite{dhn2016}, are deep hashing based methods. Neighborhood Discriminant Hashing (NDH)\cite{ndh2017}, Column Sampling Based Discrete Supervised Hashing (COSDISH) \cite{cosdish2016}, Supervised Discrete Hashing (SDH)\cite{sdh2015}, Fast supervised Hashing (FastH)\cite{fasth2014}, and Latent Factor Hashing (LFH)\cite{lfh2014} are supervised learning methods. Our method outperforms all baseline approaches. Note, that all deep hashing baseline models and our method use CNN-F as the feature extractor for a fair comparison. Furthermore, the mAP-score for the proposed DCCF method for $9$ bits, i.e., our method without an ensemble technique, is $\mathrm{mAP}_{\mathrm{9-bits}} = 0.7796$. This mAP-score is the mean of $20$ runs and the $0.95$ confidence interval for this mean is $[0.7758, 0.7835]$. Thus, we can see that even by using only $9$ bits our method outperforms the state-of-the-art methods for $12$ bits. This reveals that the gain in the performance does not come from the use of the ensemble technique alone, but our proposed feature space generation followed by the binarization could generate efficient binary codes. 
\begin{table}
\centering
\caption{\scriptsize mAP on CIFAR-10 dataset. The scores for DDSH, DSDH, DPSH, DSH, DHN, NDH, COSDISH, SDH, FastH, LFH are cited from \cite{ddsh2018}. The scores for DTSH are cited from \cite{dtsh2016}. The same experimental setup was used for all methods (see \cite{dtsh2016}). The best accuracy values are shown in boldface.}
\begin{tabular}{l l l l l l}
    \toprule      
    Method & 9 bits & 12 bits &24 bits &32 bits &48 bits\\
  \midrule 
    Ours (DCCH) & \textbf{0.7796} & \textbf{0.7936} & \textbf{0.8306} & \textbf{0.8414} & \textbf{0.8512} \\ 
    DDSH \cite{ddsh2018} & - & 0.7695 & 0.8289 & 0.8352 & 0.8194 \\
    DSDH \cite{dsdh2017} & - & 0.7442 & 0.7868 & 0.7991 & 0.8142 \\
    DTSH \cite{dtsh2016} & - & 0.7100 & 0.7500 & 0.7650 & 0.7740 \\
    DPSH \cite{dpsh2015}  & -  & 0.6844 & 0.7225 & 0.7396 & 0.7460 \\
    DSH \cite{dsh2016} & -  & 0.6457 & 0.7492 & 0.7857 & 0.8113 \\
    DHN \cite{dhn2016} & -  & 0.6725 & 0.7107 & 0.7045 & 0.7135 \\ 
    \midrule
    NDH \cite{ndh2017} & -  & 0.5620 & 0.6404 & 0.6515 & 0.6772 \\
    COSDISH \cite{cosdish2016} & - & 0.6085 & 0.6827 & 0.6959 & 0.7158 \\
    SDH \cite{sdh2015} & -  & 0.5200 & 0.6461 & 0.6577 & 0.6688 \\
    FastH \cite{fasth2014} & -  & 0.6292 & 0.6731 & 0.6870 & 0.7163 \\
    LFH \cite{lfh2014} & -  & 0.4009 & 0.6047 & 0.6571 & 0.6995 \\ %\cline{1-5}
    %ITQ \cite{itq2013} + CNN & 0.237 & 0.246 & 0.255 & 0.261 \\
    \bottomrule
\end{tabular}
\label{tab:mapCIFAR10}
\end{table}
\subsection{Experimental Results on the NUS-WIDE Dataset}
\label{nuswide-final}
%  \begin{figure}
%     \begin{subfigure}{0.25\textwidth}  
%             \vspace{-1mm}
%                         \includegraphics[scale=.3]{fig/results/CIFAR109dim.pdf}
%              \scriptsize \subcaption{CIFAR-10 dataset}           
%              \includegraphics[scale=.3]{fig/results/nuswide10dimPerson.pdf}
%              \scriptsize \subcaption{ NUS-WIDE - person category.}
%              \includegraphics[scale=.3]{fig/results/nuswide10dimAnimal.pdf}
%              \scriptsize \subcaption{ NUS-WIDE - animal category. }                    
%     \end{subfigure}%
%     \begin{subfigure}{0.25\textwidth}
%         \vspace{-1mm}
% \includegraphics[scale=.3]{fig/results/mscoco64dimCar.pdf}
%              \scriptsize \subcaption{ MSCOCO - car category.}           
%             \includegraphics[scale=.3]{fig/results/mscoco64dimPlane.pdf}
%              \scriptsize \subcaption{ MSCOCO - plane category.}
%              \includegraphics[ scale=.3]{fig/results/mscoco64dimTV.pdf}
%              \scriptsize \subcaption{MSCOCO - TV category.}  
%     \end{subfigure}
%     %\label{fig:space}
%     \caption{Two dimensional visualisation of feature space using t-SNE. }\label{fig:space}
% \end{figure}
So far, we presented the experimental result on single-labeled image dataset. In this case, our method can exploit the equivalence to LDA. For multi-labeled images, finding a suitable feature space is inherently more difficult. We developed our feature space optimization method such that it is generally applicable to image retrieval, especially multi-labeled datasets. Thus, the benchmarks for multi-labeled datasets are of particular interest. The NUS-WIDE \cite{nuswide} dataset includes $269{,}648$ multi-labeled images. Each image is labeled by $81$ concepts. As proposed in DDSH, we select the $186{,}577$ images which belong to the $10$ most frequent concepts. We randomly select $1{,}867$ images as query images. The remaining images constitute the gallery dataset. From this gallery dataset, we sample $5{,}000$ images for training our model.\par The parameters of NUS-WIDE were found by a random search and are depicted in Table \ref{tab:parameters}. The parameters remain almost unchanged in comparison to CIFAR-10. However, NUS-WIDE is much more complex due to the larger number of categories and the fact that the images are annotated by multiple labels. To capture this complexity of the data we use a larger batch size to have a better estimation of the covariance matrices. Note that we could project to a higher dimensional space to avoid the use of the ensemble technique. In particular, we could design a loss based on the sum of up to $81$ correlation coefficients for NUS-WIDE. In such a situation, at least as many categories as correlation coefficients have to be present in a batch for the estimation of the covariance matrices during the loss computation. This implies that we need a larger batch size. In general, we cannot ensure that all of these categories are present in each batch due to the limitation of the GPU memory. Therefore, we only project to a 10-dimensional space and use our proposed ensemble technique for generating longer codes.
\subsubsection{Training Loss and Learned Feature Space}
The training loss for our $10$-dimensional projections is depicted in Fig.  \ref{fig:nusloss}. We know through the bounds of the correlation coefficient that the minimal achievable loss is $-10$. The training loss comes close to this theoretical bound of $-10$.\par
%\subsection{Feature Space}
For the t-SNE visualization of multi-labeled images each data point should be highlighted by multiple colors to indicate the different labels for each image. Therefore, we present the t-SNE visualization where we highlight only some exemplary categories to evaluate if an appropriate feature space was found by our method. In Fig. \ref{fig:person}, 
we highlight the pictures which contain the category \lq person\rq. We can see that images sharing this category are close in the resulting feature space. As the number of pictures containing this category is large, it is likely that there are images which also share a more dominant category, e.g., a car in the foreground. This explains why we can see different sub-clusters and not all person images are within one big cluster. In Fig. \ref{fig:animal}, 
we highlight the images which contain the category \lq animal\rq. Again, we see a separation of the animal images to the other images. We can see sub-clusters as well. These sub-clusters can indicate images belonging to sub-categories, e.g., images depicting a dog. 

\subsubsection{Mean Average Precision}
In Table \ref{tab:mapNUSWIDE}, the mAP-scores for NUS-WIDE are summarized. For a lower number of bits, our method based on category information have a slight decrease in mAP values than the state-of-the-art methods. When calculating the mAP, images sharing multiple categories are treated as similar as images which share only one category. Our method based on category information will tend to preserve more information resulting in a feature space where images sharing more categories are closer to each other than images sharing a few categories. This property is not considered in the benchmarks. The achieved mAP-scores validate this assumption.
\begin{table}
\centering
\caption{\scriptsize mAP@5000 on NUS-WIDE dataset. The scores of the baselines are cited from \cite{ddsh2018}. The same experimental setup was used for all methods (see \cite{ddsh2018}). The best accuracy values are shown in boldface.}
\begin{tabular}{l l l l l}
    \toprule
    Method & 12 bits &24 bits &32 bits &48 bits\\
    \midrule
Ours (DCCH) & 0.7821 & 0.8143 & \textbf{0.8248} & \textbf{0.8340} \\ 
%\midrule
DDSH \cite{ddsh2018} & 0.7911 & \textbf{0.8165} & 0.8217 & 0.8259 \\
    DSDH \cite{dsdh2017} & \textbf{0.7916} & 0.8059 & 0.8063 & 0.8180 \\
    DPSH \cite{dpsh2015} & 0.7882 & 0.8085 & 0.8167 & 0.8234 \\
    DSH \cite{dsh2016} & 0.7622 & 0.7940 & 0.7968 & 0.8081 \\
    DHN \cite{dhn2016} & 0.7900 & 0.8101 & 0.8092 & 0.8180 \\ 
    \midrule
    NDH \cite{ndh2017} & 0.7015 & 0.7351 & 0.7447 & 0.7449 \\
    COSDISH \cite{cosdish2016} & 0.7303 & 0.7643 & 0.7868 & 0.7993 \\
    SDH \cite{sdh2015} & 0.7385 & 0.7616 & 0.7697 & 0.7720 \\
    FastH \cite{fasth2014} & 0.7412 & 0.7830 & 0.7948 & 0.8085 \\
    LFH \cite{lfh2014} & 0.7049 & 0.7594 & 0.7778 & 0.7936 \\
    \bottomrule
\end{tabular}
\label{tab:mapNUSWIDE}
\end{table}
\subsection{Experimental Results on the MS-COCO Dataset} \label{sec:resmscoco}
To further assess the performance of our proposed method, we use the MS-COCO dataset \cite{mscoco2014} which is a multi-labeled dataset where each image is labeled by $80$ object categories. Jian et al. \cite{ddsh2018} do not provide the results for this dataset in their DDSH paper. Hence, we follow the experimental setup of the state-of-the-art HashNet \cite{hashnet2017} method by Cao et al. We will not use the ensemble technique in this case to elaborate on the performance of our method based solely on the proposed feature extraction in combination with ITQ. According to Cao et al., we prune the images with no category information from the dataset and obtain $122{,}218$ images by combining the training and validation images. We randomly sample $5{,}000$ images for the query set. The remaining images constitute the gallery dataset. We randomly sample $10{,}000$ training images from this gallery dataset.
The parameters for the training were found by a random search and are summarized in Table \ref{tab:parameters}. To have a fair comparison to HashNet, we use the same baseline model, i.e., AlexNet instead of CNN-F in this case. Since we evaluate the performance of our model without the ensemble technique in this case, we choose the projected dimension during training to the number of bits.
\begin{figure*} 
\centering
\begin{subfigure}{.3\textwidth}
  \centering
  \vspace{0cm}
  \includegraphics[width=0.5\linewidth]{fig/appendix/mscoco/query3/query.jpg}
  \caption{\scriptsize Query image}
\end{subfigure}%
\begin{subfigure}{.7\textwidth}
 % \centering
  \vspace{0.7cm}
   %\hspace{-2cm}
  \includegraphics[width=0.32\linewidth]{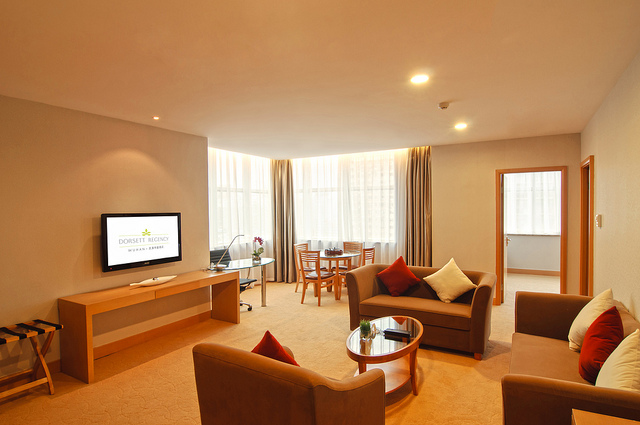}\hfil
  \includegraphics[width=0.32\linewidth]{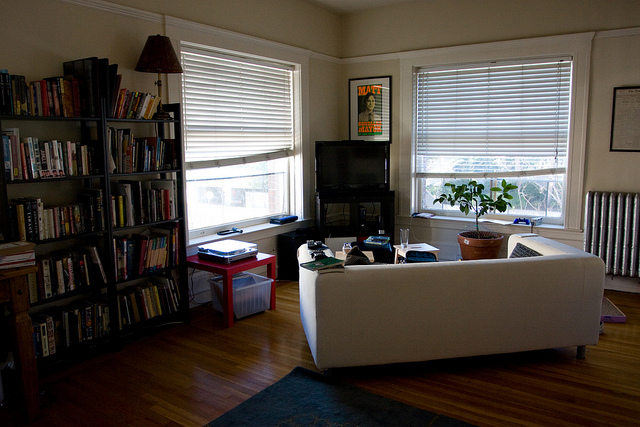}\hfil
  \includegraphics[width=0.32\linewidth]{fig/appendix/mscoco/query3/res3.jpg}
  \caption{\scriptsize Top 3 retrieved results}
  \label{fig:sampleResults64bitsMSCOCO3}
\end{subfigure}
\caption{ \scriptsize Image retrieval on MS-COCO with our 64-bits. Categories of query image: chair, potted plant, dining table, TV, book. Categories of retrieved results: $1^{st}$ - chair, couch, potted plant, dining table, TV, remote. $2^{nd}$ - couch, potted plant, dining table, TV, book. $3^{rd}$ - chair, couch, dining table, book.}
\label{fig:1}
\end{figure*}
\begin{figure*} 
\centering
\begin{subfigure}{.3\textwidth}
  \centering
  \vspace{0.4cm}
  \includegraphics[ height=0.5\linewidth, width=0.9\linewidth]{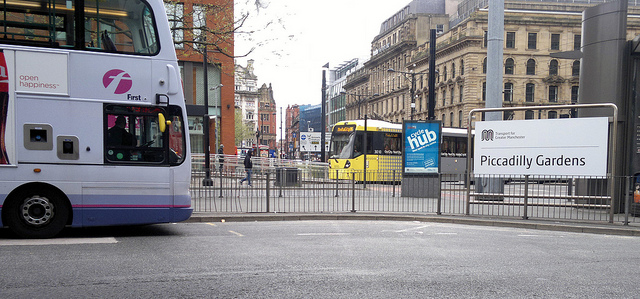}
  \caption{\scriptsize Query image}
\end{subfigure}%
\begin{subfigure}{.7\textwidth}
 % \centering
  %\vspace{0cm}
   %\hspace{-2cm}
  \colorbox{red}{\includegraphics[width=0.32\linewidth]{fig/appendix/mscoco/query2/res1.jpg}}\hfil
  \includegraphics[width=0.32\linewidth]{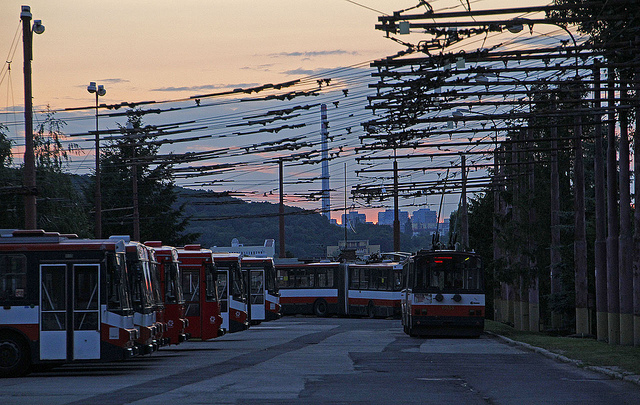}\hfil
  \includegraphics[width=0.32\linewidth]{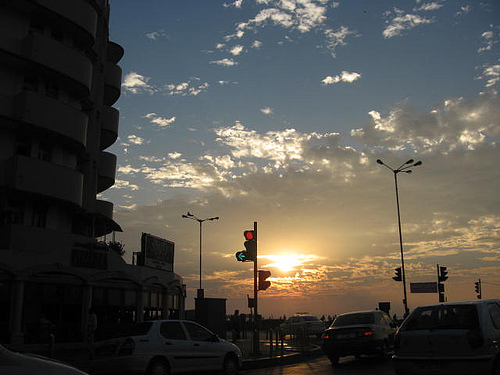}
  \caption{\scriptsize Top 3 retrieved results}
  \label{fig:2}
\end{subfigure}
\caption{ \scriptsize Image retrieval on MS-COCO with our 64-bits. Categories of query image: person, bus. Categories of retrieved results: $1^{st}$ - car, truck, traffic light. $2^{nd}$ - bus, train. $3^{rd}$ - person, car, traffic light. The $1^{st}$ retrieved image does not share any category with the query. Apparently, the road depicted in this picture led to the retrieval since a road is depicted in the query even though the images are not labeled with this concept.}
\label{fig:NUSWIDE}
\end{figure*}
\begin{figure*} 
\centering
\begin{subfigure}{.3\textwidth}
  \centering
  \vspace{.75cm}
  \includegraphics[width=0.9\linewidth]{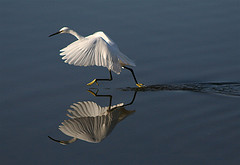}
  \caption{\scriptsize Query image}
\end{subfigure}%
\begin{subfigure}{.7\textwidth}
 % \centering
  \vspace{-.3cm}
   %\hspace{-2cm}
  \includegraphics[width=0.32\linewidth]{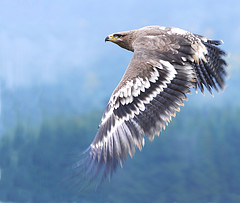}\hfil
  \includegraphics[width=0.32\linewidth, height = 0.35\textwidth ]{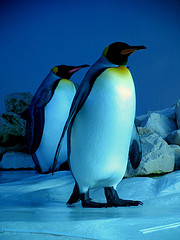}\hfil
  \includegraphics[width=0.32\linewidth]{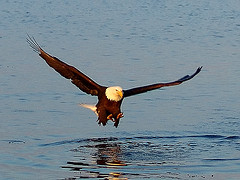}
  \caption{\scriptsize Top 3 retrieved results}
  \label{fig:sampleResults64bitsMSCOCO3}
\end{subfigure}
\caption{\scriptsize Image retrieval on NUS-WIDE with our 48-bits. Categories of
query image: water, animal, birds, reflection. Categories of retrieved results: $1^{st}$ -
sky, animal. $2^{nd}$ - animal, birds. $3^{rd}$ - water, animal, lake, birds.}
\label{fig:3}
\end{figure*}
\subsubsection{Training Loss and Learned Feature Space }
We consider the training loss for the different bit constellations $16$, $32$, $48$, and $64$. In Fig. \ref{fig:mscoco17loss}, the loss for our 16-dimensional space is depicted. We can see that the loss reaches approximately $-15$. The theoretical minimum is $-16$. In Fig. \ref{fig:mscoco33loss}, 
the loss for the $32$ dimensional space corresponding to the $32$ bit variant is plotted. Here as well, the loss is less which is expected since the theoretical minimum is $-32$ in this case. Since it is a mutli-labeled image dataset, the projections obtained by CCA cannot bring the angle between the label vectors and feature vectors exactly to zero which would have resulted in maximum correlation coefficeints. So the lower bound cannot be exactly reached in this case in contrast to the single-labeled CIFAR-10 dataset.   %\subsection{Feature Space}
Similar observations were visualized for training networks with $48$ and $64$ bit variants. For the t-SNE visualization, we consider the $64$-dimensional feature space. As MS-COCO is a multi-labeled dataset, we cannot easily present all category information within one t-SNE visualization. Therefore, we chose three sample categories to give an impression of the learned feature space. In Fig. \ref{fig:car}, 
we highlight the feature embeddings of the images which share the category information \lq car\rq. We can see that car images are in general grouped closely together. However, we also have different clusters which separate car images. The reasons for these sub-clusters are the presence of more dominant categories within these images. In Fig. \ref{fig:plane}, we can see an example of such a dominant category. The feature embeddings corresponding to plane images are marked in red. In this case, we could see a clear distinction of the corresponding cluster to the rest of the feature embeddings. If we look at Fig. \ref{fig:car}
again, we see that although there are car images in some images which also depict planes, the plane category is dominant in the feature space. The reason for this is that many images from various scenes depict cars. However, planes are usually the dominant object within the pictures due to their size. In Fig. \ref{fig:tv},
we highlight the images with the category \lq TV\rq. Again, we see an aggregation of the points in the feature space. Overall, a feature space suitable for image retrieval was found since images sharing the same categories are close together.
\subsubsection{Mean Average Precision}
In Table \ref{tab:mapMSCOCO}, the mAP-scores for $5{,}000$ retrieved images per query are summarized. As we compare to HashNet \cite{hashnet2017} approach, we adopt the same experimental setup. All deep hashing methods use AlexNet as the feature extractor instead of CNN-F in this case. HashNet, Deep Hashing Network for Efficient Similarity Retrieval (DHN (ESR)) \cite{dhnesr2016}, Deep Neural Network Hashing (DNNH) \cite{dnnh2015}, and Convolutional Neural Network Hashing (CNNH) \cite{cnnh2014} are deep hashing based supervised methods. Supervised Discrete Hashing (SDH) \cite{sdh2015}, Kernel-based Supervised Hashing (KSH) \cite{ksh2012}, and Supervised Iterative Quantization Hashing (ITQ-CCA) \cite{itq2013} are non-deep supervised hashing methods. Iterative Quantization (ITQ) \cite{itq2013}, Binary Reconstructive Embedding (BRE) \cite{bre2009}, and Spectral Hashing (SH) \cite{sh2009} are unsupervised methods and Locality Sensitive Hashing (LSH) is a data-independent method. 
For $32$, $48$, and $64$ bits our method performs better than the state-of-the-art. For 16-bits our DCCH method have comparable performance with HashNet and DHN (ESR). The gain in performance through our method comes mainly due to our effective feature extraction by DCCF and the combination with ITQ.
\begin{table}
\centering
\caption{\scriptsize mAP@5000 on MS-COCO dataset. The scores of the baselines are cited from \cite{hashnet2017}. The same experimental setup was used for all methods (see \cite{hashnet2017}). The best accuracy values are shown in boldface.}
%\resizebox{0.4\columnwidth}{!}{%
\begin{tabular}{l l l l l}
    \toprule
    Method & 16 bits & 32 bits &48 bits &64 bits\\
    \midrule
    Ours (DCCH) & 0.6592 & \textbf{0.7293} & \textbf{0.7312} & \textbf{0.7388} \\ 
    HashNet \cite{hashnet2017} & \textbf{0.6873} & 0.7184 & 0.7301 & 0.7362 \\
    DHN(ESR) \cite{dhnesr2016} & 0.6774 & 0.7013 & 0.6948 & 0.6944 \\
    DNNH \cite{dnnh2015} & 0.5932 & 0.6034 & 0.6045 & 0.6099 \\
    CNNH \cite{cnnh2014} & 0.5642 & 0.5744 & 0.5711 & 0.5671 \\ 
    \midrule
    SDH \cite{sdh2015} & 0.5545 & 0.5642 & 0.5723 & 0.5799 \\
    KSH \cite{ksh2012} & 0.5212 & 0.5343 & 0.5343 & 0.5361 \\
    ITQ-CCA \cite{itq2013} & 0.5659 & 0.5624 & 0.5297 & 0.5019 \\ 
    \midrule
    ITQ \cite{itq2013} & 0.5818 & 0.6243 & 0.6460 & 0.6574 \\
    BRE \cite{bre2009} & 0.5920 & 0.6224 & 0.6300 & 0.6336 \\
    SH \cite{sh2009} & 0.4951 & 0.5071 & 0.5099 & 0.5101 \\ 
    \midrule
    LSH \cite{lsh1999} & 0.4592 & 0.4856 & 0.5440 & 0.5849 \\
     \bottomrule
\end{tabular}%
%}
%\captionsetup{justification=centering}
\label{tab:mapMSCOCO}
\end{table}
\subsection{Example Queries}
Finally, we present a few example queries based on our proposed method.  
%We can see a bird and also water in the third retrieved image. Interestingly, from a benchmark perspective, any retrieved animal image would be sufficient since it would share at least the animal category with the query item. Not only animal images are returned, but also all returned images depict birds. This was intended by our approach as described in Section \ref{sec:multidimembeddingforpairwise}. Moreover, the first retrieved image is annotated in the NUS-WIDE dataset with \lq sky\rq\ and \lq animal\rq\ but not with \lq bird\rq\ altough the image clearly depicts a bird. First, this shows that the datasets might contain inaccurate label information. Secondly, it emphasizes the usefulness of content-based image retrieval which we discussed in Section \ref{sec:cbir}. If we would rely on the textual annotations and we would query for the images with the label \lq bird\rq\ the first retrieved image in Fig. \ref{fig:sample48bitsNUSWIDE} 
%would not be returned altough its content is very similar to the query.
To visualize an example of our DCCH method based on category labels, the results of retrieval results for $64$ bits on MS-COCO are shown in Fig. \ref{fig:1}. In Fig. \ref{fig:NUSWIDE}, we show a case where we have wrong retrieval probably because the query image contains \lq road\rq \hspace{.05cm} information but images are not labeled with that concept. We can see that our method deals with the multi-labeled query images very well. Overall, the example queries reflect the results of the mAP-scores which verify that the generated binary codes are well suited for image retrieval. Fig. \ref{fig:3} shows an example query result on the NUS-WIDE dataset. Interestingly, the retrieved images also contain images with birds flying similar to the semantic content of the query image. Retrieval results have more or less the same background information which is the blue color of sky, ice, and water for the top three retrieved images respectively.
%In the appendix, we show further example retrieval results and also present some results, where our proposed method fails, to show its limitations. Moreover, the images in the appendix show the variety of semantic concepts in the NUS-WIDE and MS-COCO datasets. However, the discussed figures in this section provide a good representation of the overall quality of our proposed method.
\section{Conclusions and Future work}
In this paper, we propose a novel approach for learning a low-dimensional feature space. We utilize the global statistics of the feature space which the pairwise and triplet based approaches fail to use. Canonical Correlation Analysis (CCA) is used to correlate the projection of feature vectors and the label vectors to
obtain a low-dimensional feature space. The visualization of training error curves prove that we are able to maximize the correlation in all directions achieving maximum class separability in all the discriminant directions which is also evident from the feature space visualizations. For single-labeld images, the obtained feature space has maximum between-class scatter and minimum within-class scatter as we utilize the equivalence between Linear Discriminant Analysis and CCA here. Once we generate a compact low-dimensional feature space, we binarize it by using the popular ITQ (Iterative Quantization) approach. Since the length of our binary codes was bounded by the number of
image categories, we proposed an ensemble network technique to generate codes of desired bit length. The measurement of mAP shows that our method significantly outperform state-of-the-art binarization schemes for image-retrieval.\par A possible future extension of our method would be to the multi-modal search. Here,
we could incorporate label information from another source, e.g., textual information. For
instance, captions in the form of sentences might be available for the training images instead
of category labels. In this case, we could find a word
embedding for representing the textual annotations as vectors. These vectors could directly
be used as the second data vector within our DCCF learning such that we correlate the output
of the CNN with this embedding. It is reasonable to assume that our proposed method would
achieve superior performance on multi-modal image retrieval.

% 
%    \item Our approach learns an optimized low-dimensional feature space for single-labeled images by utilizing the equivalence between LDA and CCA and address the problems of DeepLDA \cite{deeplda2015} approach (Section \ref{sec:FUNLDA});
%    \item We extend the optimized feature space learning to multi-labled images with CCA approach by designing appropriate category indicator matrices (Section \ref{sec:dclsinglelabeledimages}, \ref{sec:dclimultilabeledimages} );
%    \item We elaborate from a linear algebra point of view to show how the correlation maximization between the two views of CCA can be interpreted for multi-labeled images (Section \ref{sec:cca-la});
%    \item We design the loss function based on the correlation coefficients of CCA for training our network to learn the optimized feature space (Section \ref{sec:loss} );
%    \item We propose an ensemble technique for generating binary codes of desired bit length from the bits obtained from our proposed feature space followed by ITQ. (Section \ref{sec:networkensemble})
% 
% \ifCLASSOPTIONcompsoc
% 
%   % The Computer Society usually uses the plural form
% 
%   \section*{Acknowledgments}
% 
% \else
% 
%   % regular IEEE prefers the singular form
% \fi

%The authors would like to thank  Christo K Thomas, Sabarinath Mahadevan, and Maria Meyer  for helpful discussions.


\ifCLASSOPTIONcaptionsoff
\newpage
\fi
\bibliographystyle{IEEEbib}
\bibliography{references}

\begin{thebibliography}{10}

\bibitem{medicalRetrieval2004}
Henning M{\"u}ller, Nicolas Michoux, David Bandon, and Antoine Geissbuhler,
\newblock ``A review of content-based image retrieval systems in medical
  applications—clinical benefits and future directions,''
\newblock {\em Int. J Med. Inform.}, vol. 73, no. 1, pp. 1--23, 2004.

\bibitem{auto2018}
Marvin Teichmann, Michael Weber, Marius Zoellner, Roberto Cipolla, and Raquel
  Urtasun,
\newblock ``Multinet: Real-time joint semantic reasoning for autonomous
  driving,''
\newblock in {\em IEEE Int. Veh. Sym. (IV)}. IEEE, 2018, pp. 1013--1020.

\bibitem{smeulders2000}
Arnold~WM Smeulders, Marcel Worring, Simone Santini, Amarnath Gupta, and Ramesh
  Jain,
\newblock ``Content-based image retrieval at the end of the early years,''
\newblock {\em {IEEE} Trans. Pattern Anal. Mach. Intell.}, , no. 12, pp.
  1349--1380, 2000.

\bibitem{bow}
Roberto Toldo, Umberto Castellani, and Andrea Fusiello,
\newblock ``The bag of words approach for retrieval and categorization of 3d
  objects,''
\newblock {\em Vis. Comp.}, vol. 26, no. 10, pp. 1257--1268, 2010.

\bibitem{fv}
Jorge S{\'a}nchez, Florent Perronnin, Thomas Mensink, and Jakob Verbeek,
\newblock ``Image classification with the fisher vector: Theory and practice,''
\newblock {\em Int. J. Comput. Vis.}, vol. 105, no. 3, pp. 222--245, 2013.

\bibitem{alexnet2012}
Alex Krizhevsky, Ilya Sutskever, and Geoffrey~E Hinton,
\newblock ``Imagenet classification with deep convolutional neural networks,''
\newblock in {\em Adv. Neur. In. Proc. Sys.}, 2012, pp. 1097--1105.

\bibitem{babenko2014}
Artem Babenko, Anton Slesarev, Alexandr Chigorin, and Victor Lempitsky,
\newblock ``Neural codes for image retrieval,''
\newblock in {\em Euro. Conf. Comput. Vis.} Springer, 2014, pp. 584--599.

\bibitem{siftcnn2018}
Liang Zheng, Yi~Yang, and Qi~Tian,
\newblock ``Sift meets cnn: A decade survey of instance retrieval,''
\newblock {\em {IEEE} Trans. Pattern Anal. Mach. Intell.}, vol. 40, no. 5, pp.
  1224--1244, 2018.

\bibitem{dpsh2015}
Wu-Jun Li, Sheng Wang, and Wang-Cheng Kang,
\newblock ``Feature learning based deep supervised hashing with pairwise
  labels,''
\newblock {\em arXiv preprint arXiv:1511.03855}, 2015.

\bibitem{cbirsurvey2008}
Ritendra Datta, Dhiraj Joshi, Jia Li, and James~Z Wang,
\newblock ``Image retrieval: Ideas, influences, and trends of the new age,''
\newblock {\em ACM Compu. Surv.}, vol. 40, no. 2, pp. 5, 2008.

\bibitem{dtsh2016}
Xiaofang Wang, Yi~Shi, and Kris~M Kitani,
\newblock ``Deep supervised hashing with triplet labels,''
\newblock in {\em Asian Conf. Comput. Vis.} Springer, 2016, pp. 70--84.

\bibitem{ddsh2018}
Qing-Yuan Jiang, Xue Cui, and Wu-Jun Li,
\newblock ``Deep discrete supervised hashing,''
\newblock {\em IEEE Trans. Image Process.}, vol. 27, no. 12, pp. 5996--6009,
  2018.

\bibitem{surveyhash2018}
Jingdong Wang, Ting Zhang, Nicu Sebe, Heng~Tao Shen, et~al.,
\newblock ``A survey on learning to hash,''
\newblock {\em {IEEE} Trans. Pattern Anal. Mach. Intell.}, vol. 40, no. 4, pp.
  769--790, 2018.

\bibitem{hashsurvey2016}
Jun Wang, Wei Liu, Sanjiv Kumar, and Shih-Fu Chang,
\newblock ``Learning to hash for indexing big data—a survey,''
\newblock {\em Proc. IEEE}, vol. 104, no. 1, pp. 34--57, 2016.

\bibitem{sh2009}
Yair Weiss, Antonio Torralba, and Rob Fergus,
\newblock ``Spectral hashing,''
\newblock in {\em Adv. Neu. Inf. Proc. Sys.}, 2009, pp. 1753--1760.

\bibitem{itq2013}
Yunchao Gong, Svetlana Lazebnik, Albert Gordo, and Florent Perronnin,
\newblock ``Iterative quantization: A procrustean approach to learning binary
  codes for large-scale image retrieval,''
\newblock {\em {IEEE} Trans. Pattern Anal. Mach. Intell.}, vol. 35, no. 12, pp.
  2916--2929, 2013.

\bibitem{dsh2016}
Haomiao Liu, Ruiping Wang, Shiguang Shan, and Xilin Chen,
\newblock ``Deep supervised hashing for fast image retrieval,''
\newblock in {\em Proc. IEEE Conf. Comput. Vis Pattern Recognit.}, 2016, pp.
  2064--2072.

\bibitem{dhn2016}
Yue Cao, Mingsheng Long, Jianmin Wang, Han Zhu, and Qingfu Wen,
\newblock ``Deep quantization network for efficient image retrieval.,''
\newblock in {\em Ass. Adv. Art. Intell.}, 2016, pp. 3457--3463.

\bibitem{unsuperviseddeephashing}
Kevin Lin, Jiwen Lu, Chu-Song Chen, and Jie Zhou,
\newblock ``Learning compact binary descriptors with unsupervised deep neural
  networks,''
\newblock in {\em Proc. IEEE Conf. Comput. Vis Pattern Recognit.}, 2016, pp.
  1183--1192.

\bibitem{unsup}
Fumin Shen, Yan Xu, Li~Liu, Yang Yang, Zi~Huang, and Heng~Tao Shen,
\newblock ``Unsupervised deep hashing with similarity-adaptive and discrete
  optimization,''
\newblock {\em IEEE Trans. Pattern Anal. Mach. Intell.}, vol. 40, no. 12, pp.
  3034--3044, 2018.

\bibitem{unsup2}
Xianglong Liu, Yadong Mu, Danchen Zhang, Bo~Lang, and Xuelong Li,
\newblock ``Large-scale unsupervised hashing with shared structure learning,''
\newblock {\em IEEE Trans. Cyber.}, vol. 45, no. 9, pp. 1811--1822, 2014.

\bibitem{unsup3}
Jian Wang, Xin-Shun Xu, Shanqing Guo, Lizhen Cui, and Xiao-Lin Wang,
\newblock ``Linear unsupervised hashing for ann search in euclidean space,''
\newblock {\em Neurocomputing}, vol. 171, pp. 283--292, 2016.

\bibitem{hashnet2017}
Zhangjie Cao, Mingsheng Long, Jianmin Wang, and S~Yu Philip,
\newblock ``Hashnet: Deep learning to hash by continuation.,''
\newblock in {\em Int. Con. Compu. Vis.}, 2017, pp. 5609--5618.

\bibitem{fisher36lda}
R.~A. Fisher,
\newblock ``The use of multiple measurements in taxonomic problems,''
\newblock {\em Anna Eugen.}, vol. 7, pp. 179--188, 1936.

\bibitem{harold1936relations}
Harold Hotelling,
\newblock ``Relations between two sets of variates,''
\newblock {\em Biometrika}, vol. 28, no. 3/4, pp. 321--377, 1936.

\bibitem{cifar10}
Alex Krizhevsky and Geoffrey Hinton,
\newblock ``Learning multiple layers of features from tiny images,''
\newblock Tech. {R}ep., Citeseer, 2009.

\bibitem{nuswide}
Tat-Seng Chua, Jinhui Tang, Richang Hong, Haojie Li, Zhiping Luo, and Yantao
  Zheng,
\newblock ``Nus-wide: a real-world web image database from national university
  of singapore,''
\newblock in {\em Proc. ACM Int. Conf. Image Video Ret.} ACM, 2009, p.~48.

\bibitem{mscoco2014}
TY~Lin, M~Maire, S~Belongie, J~Hays, P~Perona, D~Ramanan, P~Doll{\'a}r, and
  CL~Zitnick,
\newblock ``Microsoft coco: Common objects in context. ineuropean conference on
  computer vision 2014 sep 6 (pp. 740-755),'' .

\bibitem{tsne2008}
Laurens van~der Maaten and Geoffrey Hinton,
\newblock ``Visualizing data using t-sne,''
\newblock {\em J. Mach. Learn. Res.}, vol. 9, no. Nov, pp. 2579--2605, 2008.

\bibitem{deeplda2015}
Matthias Dorfer, Rainer Kelz, and Gerhard Widmer,
\newblock ``Deep linear discriminant analysis,''
\newblock {\em arXiv preprint arXiv:1511.04707}, 2015.

\bibitem{fisherlda2005}
Max Welling,
\newblock ``Fisher linear discriminant analysis,''
\newblock {\em Depart. of Comp. Sci., Univ. Tor.}, vol. 3, no. 1, 2005.

\bibitem{mardia}
K.~V. Mardia, J.~T. Kent, and Bibby,
\newblock {\em Multivariate Analysis},
\newblock Acad. Press., 1979.

\bibitem{bartlett1938}
M.~S. Bartlett,
\newblock ``Further aspects of the theory of multiple regression,''
\newblock {\em Math. Proc. Camb. Phil. Soc.}, vol. 34, no. 1, pp. 33--40, 1938.

\bibitem{innerproduct2011}
Adam Towsley, Jonathan Pakianathan, and David~H Douglass,
\newblock ``Correlation angles and inner products: Application to a problem
  from physics,''
\newblock {\em ISRN App. Math.}, vol. 2011, 2011.

\bibitem{andrew2013}
Galen Andrew, Raman Arora, Jeff Bilmes, and Karen Livescu,
\newblock ``Deep canonical correlation analysis,''
\newblock in {\em Int. Conf. Mach. Learn.}, 2013, pp. 1247--1255.

\bibitem{hardoon2004}
David~R Hardoon, Sandor Szedmak, and John Shawe-Taylor,
\newblock ``Canonical correlation analysis: An overview with application to
  learning methods,''
\newblock {\em Neural Comp.}, vol. 16, no. 12, pp. 2639--2664, 2004.

\bibitem{cnnf2014}
Ken Chatfield, Karen Simonyan, Andrea Vedaldi, and Andrew Zisserman,
\newblock ``Return of the devil in the details: Delving deep into convolutional
  nets,''
\newblock {\em arXiv preprint arXiv:1405.3531}, 2014.

\bibitem{jegou2010}
Herv{\'e} J{\'e}gou, Matthijs Douze, Cordelia Schmid, and Patrick P{\'e}rez,
\newblock ``Aggregating local descriptors into a compact image
  representation,''
\newblock in {\em Proc. IEEE Conf. Comput. Vis Pattern Recognit.} IEEE, 2010,
  pp. 3304--3311.

\bibitem{networkensembleclassification}
Giorgio Giacinto and Fabio Roli,
\newblock ``Design of effective neural network ensembles for image
  classification purposes,''
\newblock {\em Image and Vision Comput.}, vol. 19, no. 9-10, pp. 699--707,
  2001.

\bibitem{ensembleImageRetrievalAverageFeatures2016}
Hsin-Kai Huang, Chien-Fang Chiu, Chien-Hao Kuo, Yu-Chi Wu, Narisa~NY Chu, and
  Pao-Chi Chang,
\newblock ``Mixture of deep cnn-based ensemble model for image retrieval,''
\newblock in {\em IEEE 5th Glob. Conf. Cons. Elec.} IEEE, 2016, pp. 1--2.

\bibitem{orb2011}
Ethan Rublee, Vincent Rabaud, Kurt Konolige, and Gary Bradski,
\newblock ``Orb: An efficient alternative to sift or surf,''
\newblock 2011.

\bibitem{grid}
Matthias Feurer and Frank Hutter,
\newblock ``Hyperparameter optimization,''
\newblock {\em Auto. Mach. Learn.}, p.~1, 2019.

\bibitem{howshouldwe2017}
Alexandre Sablayrolles, Matthijs Douze, Nicolas Usunier, and Herv{\'e}
  J{\'e}gou,
\newblock ``How should we evaluate supervised hashing?,''
\newblock in {\em IEEE Int. Conf. Aco. Spe. Sig. Proc.} IEEE, 2017, pp.
  1732--1736.

\bibitem{dsdh2017}
Qi~Li, Zhenan Sun, Ran He, and Tieniu Tan,
\newblock ``Deep supervised discrete hashing,''
\newblock in {\em Adv. Neur. In. Proc. Sys.}, 2017, pp. 2482--2491.

\bibitem{ndh2017}
Dekui Ma, Jian Liang, Ran He, and Xiangwei Kong,
\newblock ``Nonlinear discrete cross-modal hashing for visual-textual data,''
\newblock {\em IEEE Multi.}, , no. 2, pp. 56--65, 2017.

\bibitem{cosdish2016}
Wang-Cheng Kang, Wu-Jun Li, and Zhi-Hua Zhou,
\newblock ``Column sampling based discrete supervised hashing.,''
\newblock in {\em Ass. Adv. Art. Intell.}, 2016, pp. 1230--1236.

\bibitem{sdh2015}
Fumin Shen, Chunhua Shen, Wei Liu, and Heng Tao~Shen,
\newblock ``Supervised discrete hashing,''
\newblock in {\em Proc. IEEE Conf. Comput. Vis Pattern Recognit.}, 2015, pp.
  37--45.

\bibitem{fasth2014}
Guosheng Lin, Chunhua Shen, Qinfeng Shi, Anton Van~den Hengel, and David Suter,
\newblock ``Fast supervised hashing with decision trees for high-dimensional
  data,''
\newblock in {\em Proc. IEEE Conf. Comput. Vis Pattern Recognit.}, 2014, pp.
  1963--1970.

\bibitem{lfh2014}
Peichao Zhang, Wei Zhang, Wu-Jun Li, and Minyi Guo,
\newblock ``Supervised hashing with latent factor models,''
\newblock in {\em Proc. 37th Int. ACM SIGIR Conf. Res. \& Devel. Infor. Ret.}
  ACM, 2014, pp. 173--182.

\bibitem{dhnesr2016}
Han Zhu, Mingsheng Long, Jianmin Wang, and Yue Cao,
\newblock ``Deep hashing network for efficient similarity retrieval.,''
\newblock in {\em Ass. Adv. Art. Intell.}, 2016, pp. 2415--2421.

\bibitem{dnnh2015}
Hanjiang Lai, Yan Pan, Ye~Liu, and Shuicheng Yan,
\newblock ``Simultaneous feature learning and hash coding with deep neural
  networks,''
\newblock in {\em Proc. IEEE Conf. Comput. Vis Pattern Recognit.}, 2015, pp.
  3270--3278.

\bibitem{cnnh2014}
Rongkai Xia, Yan Pan, Hanjiang Lai, Cong Liu, and Shuicheng Yan,
\newblock ``Supervised hashing for image retrieval via image representation
  learning.,''
\newblock in {\em Ass. Adv. Art. Intell.}, 2014, vol.~1, p.~2.

\bibitem{ksh2012}
Wei Liu, Jun Wang, Rongrong Ji, Yu-Gang Jiang, and Shih-Fu Chang,
\newblock ``Supervised hashing with kernels,''
\newblock in {\em Proc. IEEE Conf. Comput. Vis Pattern Recognit.} IEEE, 2012,
  pp. 2074--2081.

\bibitem{bre2009}
Brian Kulis and Trevor Darrell,
\newblock ``Learning to hash with binary reconstructive embeddings,''
\newblock in {\em Adv. Neu. Info. Proc. Sys.}, 2009, pp. 1042--1050.

\bibitem{lsh1999}
Aristides Gionis, Piotr Indyk, Rajeev Motwani, et~al.,
\newblock ``Similarity search in high dimensions via hashing,''
\newblock in {\em Vldb}, 1999, vol.~99, pp. 518--529.

\end{thebibliography}
\vspace{-2.5cm}
\begin{IEEEbiography}[{\includegraphics [width=1in,height=1.3in,] {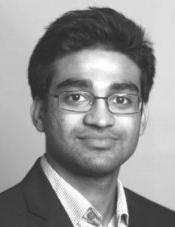}}]{Abin Jose}
received the bachelors degree in Electronics and Communication Engineering from College of Engineering, Trivandrum, India, in 2010. He worked as a software engineer in Motorola Solutions India for 2 years and as a research student in Spectrum Lab at Indian Institute of Science, Bangalore for 1 year. He received the MSc degree in Communications Engineering from the RWTH University, Aachen, in 2016. He is currently pursuing the PhD degree in the Institut f\"{u}r Nachrichtentechnik, RWTH University, Aachen, Germany. His research interests are computer vision and machine learning.
\end{IEEEbiography}
\vspace{-2.5cm}
\begin{IEEEbiography}[{\includegraphics[width=1in,height=1.3in,clip]{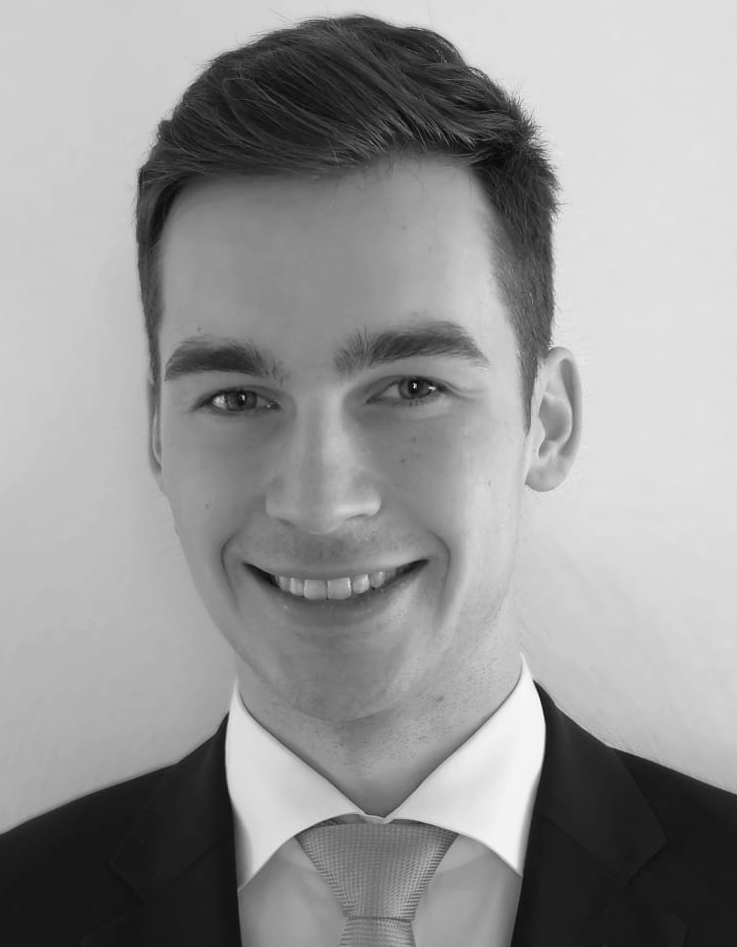}}]{Erik Stefan Ottlik}
was born in Aachen, Germany in 1994. He received the Master’s degree in Electrical Engineering, Information Technology and Computer Engineering from RWTH Aachen University, Germany, in 2019. His research interests include computer vision in the field of autonomous driving and corresponding machine learning techniques.
\end{IEEEbiography}
\vspace{-2.5cm}
\begin{IEEEbiography}[{\includegraphics[width=1in,height=1.3in,clip]{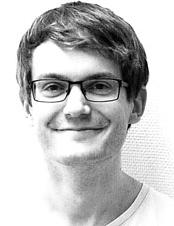}}]{Christian Rohlfing}
received the diploma degree in Computer Engineering, focus on Media Engineering, from RWTH Aachen University in 2012 and his PhD in Electrical Engineering in 2018. He is currently chief engineer at the Institut f\"{u}r Nachrichtentechnik, RWTH Aachen University. His research interests are bio and audio signal processing and machine learning.
\end{IEEEbiography}
\vspace{-2.5cm}
\begin{IEEEbiography}[{\includegraphics[width=1in,height=1.3in]{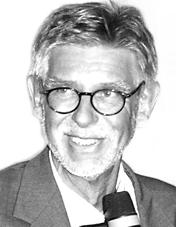}}]{Jens-Rainer Ohm}
received the Dipl.-Ing.,
Dr.Ing., and Habilitation degrees from Technical
University of Berlin, Berlin, Germany, in 1985,
1990, and 1997, respectively.
He has been participating in the Moving Picture Experts Group (MPEG) since 1998. He has
been a Full Professor and the Chair of Institute of
Communication Engineering with RWTH Aachen
University, Aachen, Germany, since 2000. He has
authored numerous papers and German and English
textbooks in multimedia signal processing, analysis,
and coding, and basics of communication engineering and signal transmission. His research interests include motion-compensated, stereoscopic and
3D image processing, multimedia signal coding, transmission and content
description, audio signal analysis, and fundamental topics of signal processing
and digital communication systems.
Dr. Ohm has been the Chair and Co-Chair of various standardization
activities in video coding, namely, the MPEG Video Subgroup since 2002, the
Joint Video Team of MPEG and ITU-T SG 16 Video Coding Experts Group
from 2005 to 2009, and currently the Joint Collaborative Team on Video
Coding (JCT-VC) and the Joint Collaborative Team on 3D Video Coding
Extensions (JCT-3V).
\end{IEEEbiography}
% \vfill
% \enlargethispage{-5in}
%In Fig. \ref{fig:beforeITQ}, we can see some 2-dimensional toy data. Different colors indicate different classes. If we apply a direct binarization by a sign-thresholding for the two dimensions, we will get bad binary codes, e.g., red and green points will be mapped to the same binary code \lq 11\rq. By using ITQ, we can find an optimal rotation matrix. After applying this rotation, we end up with Fig. \ref{fig:afterITQ}. Then, a simple sign-thresholding would reveal the desired binary codes such that different classes are mapped to different codes. Note that we also include improper rotations here, which flip the axes even though they are strictly speaking not rotation matrices as their determinant is $-1$.
\end{document}